\documentclass[twocolumn,showpacs,aps,prd]{revtex4}
\usepackage{graphicx}
\usepackage{epsf}  

\newcommand{\BaBarYear}    {04}
\newcommand{\BaBarNumber}  {008}
\newcommand{\SLACPubNumber} {10381}
\newcommand{\LANLNumber} {0403025}

\input pubboard/babarsym

\RequirePackage{xspace}

\newcommand{\calH}{\ensuremath{{\cal H}}}
\newcommand{\pvec}{{\bf p}}
\newcommand{\acp}{\ensuremath{\calA_{ch}}}
\newcommand{\calB}{\ensuremath{{\cal B}}}

\newcommand{\bfemsix}{${\cal B}(10^{-6}$)}

\newcommand{\DE}{\ensuremath{\Delta E}}

\newcommand{\xf}{\ensuremath{{\cal F}}}
\newcommand{\hel}{\ensuremath{{\cal H}}}
\newcommand{\thetaT}{\ensuremath{\theta_{\rm T}}}
\newcommand{\costhr}{\ensuremath{\cos\thetaT}}

\newcommand\etal{{\it et al.}}
\newcommand{\half}{\ensuremath{{1\over2}}}

\newcommand{\bma}[1]{\boldmath{$#1$}}
\newcommand{\msp}{\phantom{$-$}}
\newcommand{\mdsp}{\phantom{--}}

\newcommand{\bfig}{\begin{figure}[htbpc!]}
\newcommand{\efig}{\end{figure}}
\newcommand\bef{\begin{figure}}
\newcommand\edf{\end{figure}}
\newcommand\dbline{\noalign{\vskip 0.10truecm\hrule}\noalign{\vskip 2pt}\noalign{\hrule\vskip 0.10truecm}}

\newcommand\sgline{\noalign{\vskip 0.10truecm\hrule\vskip 0.10truecm}}
\newcommand\beq{\begin{equation}}
\newcommand\eeq{\end{equation}}
\newcommand\bear{\begin{array}}
\newcommand\enar{\end{array}}
\newcommand\beqa{\begin{eqnarray}}
\newcommand\eeqa{\end{eqnarray}}
\newcommand\ben{\begin{enumerate}}
\newcommand\een{\end{enumerate}}

\newcommand{\UfourS}{\ensuremath{\Upsilon(4S)}}

\newcommand{\etappp}{\ensuremath{\eta_{3\pi}}}
\newcommand{\etatogg}{\ensuremath{\eta\ra\gaga}}
\newcommand{\etatoppp}{\ensuremath{\eta\ra\pi^+\pi^-\pi^0}}

\newcommand{\etapepp}{\ensuremath{\etapr_{\eta\pi\pi}}}
\newcommand{\etaptoepp}{\ensuremath{\etapr\ra\eta\pip\pim}}
\newcommand{\etaprg}{\ensuremath{\etapr_{\rho\gamma}}}
\newcommand{\etaptorg}{\ensuremath{\etapr\ra\rho^0\gamma}}

\newcommand{\etaprp}{\ensuremath{\eta^{(\prime)}}}		% ``eta prime, parenth''

\newcommand{\om}{\ensuremath{{\omega}}}

\newcommand{\omtoppp}{\ensuremath{{\omega\ra\pip\pim\piz}}}

\newcommand{\phitoKpKm}{\ensuremath{\phi\ra\Kp\Km}}

\newcommand{\Kst}{\ensuremath{K^*}}
\newcommand{\Kstp}{\ensuremath{\Kstarp}}
\newcommand{\Kstz}{\ensuremath{\Kstarz}}
   \newcommand{\KstpKppiz}{\ensuremath{\Kstarp_{K^+\pi^0}}}
   
   \newcommand{\KstpKspip}{\ensuremath{\Kstarp_{\Kz\pi^+}}}
   
   \newcommand{\KstzKppim}{\ensuremath{\Kstarz_{K^+\pi^-}}}

   \newcommand{\rhop}{\ensuremath{\rho^+}}
   
   \newcommand{\rhom}{\ensuremath{\rho^-}}
   
   \newcommand{\rhoz}{\ensuremath{\rho^0}}

\newcommand{\kzs}{\ensuremath{\KS}}
\newcommand{\thrpi}{\ensuremath{\pi^+\pi^-\pi^0}}

\newcommand{\tpi}{\ensuremath{3\pi}}
\newcommand{\epp}{\ensuremath{\eta\pi\pi}}
\newcommand{\fivepi}{\ensuremath{\eta_{3\pi}\pi\pi}}
\newcommand{\rg}{\ensuremath{\rho\gamma}}

\newcommand{\Dcontrol}{\ensuremath{\Bm\ra\pim\Dz~{\rm and}~\Bm\ra\rho^-\Dz~{\rm with}~\Dz\ra K^{-}\pip\piz}}

\newcommand{\fetaK}{\ensuremath{\eta K}}
\newcommand{\etaK}{\ensuremath{\B\ra\fetaK}}

\newcommand{\fetaKp}{\ensuremath{\eta K^+}}
\newcommand{\etaKp}{\ensuremath{\Bp\ra\fetaKp}}

\newcommand{\fetapiz}{\ensuremath{\eta\piz}\xspace}
\newcommand{\etapiz}{\ensuremath{\Bz\ra\fetapiz}\xspace}
\newcommand{\Betapiz}{\ensuremath{\calB(\etapiz)}\xspace}
\newcommand{\retapiz}{\ensuremath{xx^{+xx}_{-xx}\pm xx}\xspace}
\newcommand{\Retapiz}{\ensuremath{(\retapiz)\times 10^{-6}}\xspace}
\newcommand{\uletapiz}{\ensuremath{xx}\xspace}
\newcommand{\ULetapiz}{\ensuremath{\uletapiz\times 10^{-6}}\xspace}
\newcommand{\setapiz}{\ensuremath{xx}\xspace}
   \newcommand{\fetaggpiz}{\ensuremath{\eta_{\gaga} \piz}\xspace}
   
   \newcommand{\fetappppiz}{\ensuremath{\eta_{3\pi} \piz}\xspace}

\newcommand{\fetaKst}{\ensuremath{\eta K^{*}}}
\newcommand{\etaKst}{\ensuremath{\B\ra\fetaKst}}

\newcommand{\fetaKstp}{\ensuremath{\eta K^{*+}}}
\newcommand{\etaKstp}{\ensuremath{\Bp\ra\fetaKstp}}
\newcommand{\BetaKstp}{\ensuremath{\calB(\etaKstp)}}
\newcommand{\retaKstp}{\ensuremath{xx^{+xx}_{-xx}\pm xx}}
\newcommand{\RetaKstp}{\ensuremath{(\retaKstp\,)\times 10^{-6}}}
\newcommand{\AetaKstp}{\ensuremath{xx\pm xx\pm xx}}
\newcommand{\setaKstp}{\ensuremath{xx}}
  \newcommand{\fetaggKstpKspip}{\ensuremath{\eta_{\gamma\gamma}K^{*+}_{\Kz \pi^+}}}
  
  \newcommand{\fetaggKstpKppiz}{\ensuremath{\eta_{\gamma\gamma} K^{*+}_{K^+\pi^0}}}
  
  \newcommand{\fetapppKstpKspip}{\ensuremath{\eta_{3\pi}K^{*+}_{\Kz \pi^+}}}
  
  \newcommand{\fetapppKstpKppiz}{\ensuremath{\eta_{3\pi}K^{*+}_{K^+\pi^0}}}

\newcommand{\fetaKstz}{\ensuremath{\eta K^{*0}}}
\newcommand{\etaKstz}{\ensuremath{\Bz\ra\fetaKstz}}
\newcommand{\BetaKstz}{\ensuremath{\calB(\etaKstz)}}
\newcommand{\retaKstz}{\ensuremath{xx^{+xx}_{-xx}\pm xx}}
\newcommand{\RetaKstz}{\ensuremath{(\retaKstz)\times 10^{-6}}}
\newcommand{\AetaKstz}{\ensuremath{xx\pm xx \pm xx}}
\newcommand{\setaKstz}{\ensuremath{xx}}
  \newcommand{\fetaggKstz}{\ensuremath{\eta_{\gamma\gamma}K^{*0}}}

  \newcommand{\fetapppKstz}{\ensuremath{\eta_{3\pi} K^{*0}}}

\newcommand{\fetarho}{\ensuremath{\eta\rho}}
\newcommand{\etarho}{\ensuremath{B\ra\fetarho}}

\newcommand{\fetarhop}{\ensuremath{\eta\rho^+}}
\newcommand{\etarhop}{\ensuremath{\Bp\ra\fetarhop}}
\newcommand{\Betarhop}{\ensuremath{\calB(\etarhop)}}
\newcommand{\retarhop}{\ensuremath{xx^{+xx}_{-xx}\pm xx}}
\newcommand{\Retarhop}{\ensuremath{(\retarhop)\times 10^{-6}}}
\newcommand{\uletarhop}{\ensuremath{xx}}
\newcommand{\ULetarhop}{\ensuremath{\uletarhop\times 10^{-6}}}

\newcommand{\setarhop}{\ensuremath{xx}}
  \newcommand{\fetaggrhop}{\ensuremath{\eta_{\gamma\gamma} \rho^+}}
  
  \newcommand{\fetappprhop}{\ensuremath{\eta_{3\pi} \rho^+}}

\newcommand{\fetarhoz}{\ensuremath{\eta\rho^0}}
\newcommand{\etarhoz}{\ensuremath{\Bz\ra\fetarhoz}}
\newcommand{\Betarhoz}{\ensuremath{\calB(\etarhoz)}}
\newcommand{\retarhoz}{\ensuremath{xx^{+xx}_{-xx}\pm xx}}
\newcommand{\Retarhoz}{\ensuremath{(\retarhoz)\times 10^{-6}}}
\newcommand{\uletarhoz}{\ensuremath{xx}}
\newcommand{\ULetarhoz}{\ensuremath{\uletarhoz\times 10^{-6}}}

\newcommand{\setarhoz}{\ensuremath{xx}}
  \newcommand{\fetaggrhoz}{\ensuremath{\eta_{\gamma\gamma} \rho^0}}
  
  \newcommand{\fetappprhoz}{\ensuremath{\eta_{3\pi} \rho^0}}

\newcommand{\fetapK}{\ensuremath{\etapr K}}
\newcommand{\etapK}{\ensuremath{\B\ra\fetapK}}

   \newcommand{\fetaprgKp}{\ensuremath{\etapr_{\rho\gamma} K^+}}
   \newcommand{\etaprgKp}{\ensuremath{\Bp\ra\fetaprgKp}}

\newcommand{\fetappiz}{\ensuremath{\etapr\piz}\xspace}
\newcommand{\etappiz}{\ensuremath{\Bz\ra\fetappiz}\xspace}
\newcommand{\Betappiz}{\ensuremath{\calB(\Bz\ra\etapr \piz)}\xspace}
\newcommand{\retappiz}{\ensuremath{xx^{+xx}_{-xx} \pm xx}\xspace}
\newcommand{\Retappiz}{\ensuremath{(\retappiz)\times 10^{-6}}\xspace}
\newcommand{\uletappiz}{\ensuremath{xx}\xspace}
\newcommand{\ULetappiz}{\ensuremath{\uletappiz\times 10^{-6}}\xspace}

\newcommand{\setappiz}{\ensuremath{xx}\xspace}
   \newcommand{\fetapepppiz}{\ensuremath{\etapr_{\eta\pi\pi} \piz}\xspace}
   
   \newcommand{\fetaprgpiz}{\ensuremath{\etapr_{\rho\gamma} \piz}\xspace}
   \newcommand{\etaprgpiz}{\ensuremath{\Bz\ra\fetaprgpiz}\xspace}

\newcommand{\fetapKst}{\ensuremath{\etapr K^{*}}}
\newcommand{\fetaprgKst}{\ensuremath{\etapr_{\rho\gamma}\Kst}}
\newcommand{\etapKst}{\ensuremath{\B\ra\etapr K^{*}}}

\newcommand{\fetapKstp}{\ensuremath{\etapr K^{*+}}}
\newcommand{\etapKstp}{\ensuremath{\Bp\ra\etapr K^{*+}}}
\newcommand{\BetapKstp}{\ensuremath{\calB(\etapKstp)}}
\newcommand{\retapKstp}{\ensuremath{xx^{+xx}_{-xx}\pm xx}}
\newcommand{\RetapKstp}{\ensuremath{(\retapKstp)\times 10^{-6}}}
\newcommand{\uletapKstp}{\ensuremath{xx}}
\newcommand{\ULetapKstp}{\ensuremath{\uletapKstp\times 10^{-6}}}
\newcommand{\setapKstp}{\ensuremath{xx}}
   \newcommand{\fetapeppKstpKspip}{\ensuremath{\etapr_{\eta\pi\pi}\Kstarp_{\Kz\pip}}}
   
   \newcommand{\fetapeppKstpKppiz}{\ensuremath{\etapr_{\eta\pi\pi}\Kstarp_{\Kp\piz}}}
   
   \newcommand{\fetaprgKstpKspip}{\ensuremath{\etapr_{\rho\gamma}\Kstarp_{\Kz\pip}}}
   
   \newcommand{\fetaprgKstpKppiz}{\ensuremath{\etapr_{\rho\gamma}\Kstarp_{\Kp\piz}}}

\newcommand{\fetapKstz}{\ensuremath{\etapr K^{*0}}}
\newcommand{\etapKstz}{\ensuremath{\Bz\ra\fetapKstz}}
\newcommand{\BetapKstz}{\ensuremath{\calB(\etapKstz)}}
\newcommand{\retapKstz}{\ensuremath{xx^{+xx}_{-xx}\pm xx}}
\newcommand{\RetapKstz}{\ensuremath{(\retapKstz)\times 10^{-6}}}
\newcommand{\uletapKstz}{\ensuremath{xx}}
\newcommand{\ULetapKstz}{\ensuremath{\uletapKstz\times 10^{-6}}}
\newcommand{\setapKstz}{\ensuremath{xx}}
   \newcommand{\fetapeppggKstz}{\ensuremath{\etapr_{\eta(\gamma\gamma)\pi\pi} \Kstarz}}
   
   \newcommand{\fetapfivepiKstz}{\ensuremath{\etapr_{\eta(3\pi)\pi\pi} \Kstarz}}
   
   \newcommand{\fetaprgKstz}{\ensuremath{\etapr_{\rho\gamma} K^{*0}}}

\newcommand{\fetaprho}{\ensuremath{\etapr\rho}}
\newcommand{\fetaprgrho}{\ensuremath{\etapr_{\rho\gamma}\rho}}
\newcommand{\etaprho}{\ensuremath{\B\ra\fetaprho}}

\newcommand{\fetaprhop}{\ensuremath{\etapr\rho^+}}
\newcommand{\etaprhop}{\ensuremath{\Bp\ra\fetaprhop}}
\newcommand{\Betaprhop}{\ensuremath{\calB(\etaprhop)}}
\newcommand{\retaprhop}{\ensuremath{xx^{+xx}_{-xx}\pm xx}}
\newcommand{\Retaprhop}{\ensuremath{(\retaprhop)\times 10^{-6}}}
\newcommand{\uletaprhop}{\ensuremath{xx}}
\newcommand{\ULetaprhop}{\ensuremath{\uletaprhop\times 10^{-6}}}

\newcommand{\setaprhop}{\ensuremath{xx}}
  \newcommand{\fetaprgrhop}{\ensuremath{\etapr_{\rho\gamma}\rhop}}
  \newcommand{\fetapepprhop}{\ensuremath{\etapr_{\eta\pi\pi}\rhop}}

\newcommand{\fetaprhoz}{\ensuremath{\etapr\rho^0}}
\newcommand{\etaprhoz}{\ensuremath{\Bz\ra\fetaprhoz}}
\newcommand{\Betaprhoz}{\ensuremath{\calB(\etaprhoz)}}
\newcommand{\retaprhoz}{\ensuremath{xx^{+xx}_{-xx}\pm xx}}
\newcommand{\Retaprhoz}{\ensuremath{(\retaprhoz)\times 10^{-6}}}
\newcommand{\uletaprhoz}{\ensuremath{xx}}
\newcommand{\ULetaprhoz}{\ensuremath{\uletaprhoz\times 10^{-6}}}

\newcommand{\setaprhoz}{\ensuremath{xx}}
\newcommand{\fomegapiz}{\ensuremath{\omega\pi^0}\xspace}
\newcommand{\omegapiz}{\ensuremath{\Bz\ra\fomegapiz}\xspace}
\newcommand{\Bomegapiz}{\ensuremath{\calB(\omegapiz)}\xspace}
\newcommand{\romegapiz}{\ensuremath{xx\pm xx \pm xx}\xspace}
\newcommand{\Romegapiz}{\ensuremath{(\romegapiz)\times 10^{-6}}\xspace}
\newcommand{\ulomegapiz}{\ensuremath{xx}\xspace}
\newcommand{\ULomegapiz}{\ensuremath{\ulomegapiz\times 10^{-6}}\xspace}

\newcommand{\fphipiz}{\ensuremath{\phi\pi^0}\xspace}
\newcommand{\phipiz}{\ensuremath{\Bz\ra\fphipiz}\xspace}
\newcommand{\Bphipiz}{\ensuremath{\calB(\phipiz)}\xspace}
\newcommand{\rphipiz}{\ensuremath{xx\pm xx \pm xx}\xspace}
\newcommand{\Rphipiz}{\ensuremath{(\rphipiz)\times 10^{-6}}\xspace}
\newcommand{\ulphipiz}{\ensuremath{xx}\xspace}
\newcommand{\ULphipiz}{\ensuremath{\ulphipiz\times 10^{-6}}\xspace}
\newcommand{\sphipiz}{\ensuremath{xx}\xspace}

\renewcommand{\retapiz}{\ensuremath{0.7^{+1.1}_{-0.9}\pm 0.3}}
\renewcommand{\uletapiz}{\ensuremath{2.5}}
\renewcommand{\setapiz}{\ensuremath{0.8}}			

\renewcommand{\retaKstp}{\ensuremath{25.6\pm4.0\pm2.4}}
\renewcommand{\AetaKstp}{\ensuremath{+0.13\pm0.14\pm0.02}}
\renewcommand{\setaKstp}{\ensuremath{9}}

\renewcommand{\retaKstz}{\ensuremath{18.6\pm2.3\pm 1.2}}
\renewcommand{\AetaKstz}{\ensuremath{+0.02\pm0.11\pm0.02}}
\renewcommand{\setaKstz}{\ensuremath{11}}

\renewcommand{\retarhop}{\ensuremath{9.2\pm3.4\pm1.0}}
\renewcommand{\uletarhop}{\ensuremath{14}}
\renewcommand{\setarhop}{\ensuremath{3.5}}

\renewcommand{\retarhoz}{\ensuremath{-1.1^{+0.7}_{-0.9}\pm0.4}}      
\renewcommand{\uletarhoz}{\ensuremath{1.5}}
\renewcommand{\setarhoz}{\ensuremath{-}}                        

\renewcommand{\retappiz}{\ensuremath{1.0^{+1.4}_{-1.0} \pm 0.8}}
\renewcommand{\uletappiz}{\ensuremath{3.7}}
\renewcommand{\setappiz}{\ensuremath{0.7}}

\renewcommand{\retapKstp}{\ensuremath{6.3^{+4.6}_{-3.6}\pm 1.8}}
\renewcommand{\uletapKstp}{\ensuremath{14}}
\renewcommand{\setapKstp}{\ensuremath{1.9}}

\renewcommand{\retapKstz}{\ensuremath{4.1^{+2.1}_{-1.8}\pm 1.2}}
\renewcommand{\uletapKstz}{\ensuremath{7.6}}
\renewcommand{\setapKstz}{\ensuremath{2.1}}

\renewcommand{\retaprhop}{\ensuremath{12.9^{+6.2}_{-5.5}\pm 2.0}}
\renewcommand{\uletaprhop}{\ensuremath{22}}
\renewcommand{\setaprhop}{\ensuremath{2.6}}

\renewcommand{\retaprhoz}{\ensuremath{0.8^{+1.7}_{-1.2}\pm 0.9}}      
\renewcommand{\uletaprhoz}{\ensuremath{4.3}}
\renewcommand{\setaprhoz}{\ensuremath{0.5}}                        

\renewcommand{\romegapiz}{\ensuremath{-0.6^{+0.7}_{-0.5}\pm0.2}}
\renewcommand{\ulomegapiz}{\ensuremath{1.2}}

\renewcommand{\rphipiz}{\ensuremath{0.2^{+0.4}_{-0.3}\pm0.1}}
\renewcommand{\ulphipiz}{\ensuremath{1.0}}
\renewcommand{\sphipiz}{\ensuremath{0.7}}

\long\def\inst#1{\par\nobreak\kern 4pt\nobreak
    {\it #1}\par\vskip 10pt plus 3pt minus 3pt}

\begin{document}
\preprint{\babar-PUB-\BaBarYear/\BaBarNumber} 
\preprint{SLAC-PUB-\SLACPubNumber} 

\begin{flushleft}
\babar-PUB-\BaBarYear/\BaBarNumber \\
SLAC-PUB-\SLACPubNumber \\
hep-ex/\LANLNumber \\
\end{flushleft}

\title{ 
\Large \bf\boldmath 
\B\ meson decays to $\etaprp K^*$, $\etaprp\rho$,
$\etaprp\pi^0$, $\omega\pi^0$, and $\phi\pi^0$ }

%% author list as of 02-Feb-2004 (597 authors)
%
\author{B.~Aubert}
\author{R.~Barate}
\author{D.~Boutigny}
\author{F.~Couderc}
\author{J.-M.~Gaillard}
\author{A.~Hicheur}
\author{Y.~Karyotakis}
\author{J.~P.~Lees}
\author{V.~Tisserand}
\author{A.~Zghiche}
\affiliation{Laboratoire de Physique des Particules, F-74941 Annecy-le-Vieux, France }
% ...
\collaboration{The \babar\ Collaboration}
\noaffiliation

\begin{abstract}
We present measurements of the branching fractions and charge
asymmetries (where appropriate) of two-body \B\ decays to $\etaprp K^*$, 
$\etaprp\rho$, $\etaprp\pi^0$, $\omega\pi^0$, and $\phi\pi^0$.
The data were recorded with the \babar\ detector at PEP-II and
correspond to $89\times 10^6$ \BB\ pairs produced in \epem\ annihilation 
through the \UfourS\ resonance.  We find significant signals for two
decay modes and measure the branching fractions $\BetaKstp = \RetaKstp$ 
and $\BetaKstz = \RetaKstz$, where the first error is statistical and the 
second systematic.  We also find evidence with significance
$\setarhop\sigma$ for a third decay mode and measure $\Betarhop = \Retarhop$.
For other channels, we set $90\%$ C.L. upper limits of 
$\Betarhoz < \ULetarhoz$,
$\BetapKstp < \ULetapKstp$,
$\BetapKstz < \ULetapKstz$, 
$\Betaprhop < \ULetaprhop$,
$\Betaprhoz < \ULetaprhoz$,
$\Betapiz < \ULetapiz$,
$\Betappiz < \ULetappiz$,
$\Bomegapiz < \ULomegapiz$,
and $\Bphipiz < \ULphipiz$.
For self-flavor-tagging modes with significant signals, the time-integrated 
charge asymmetries are
$\acp(\fetaKstp)=\AetaKstp$ and $\acp(\fetaKstz)=\AetaKstz$.
\end{abstract}


\maketitle

\newpage

\setcounter{footnote}{0}

\section{Introduction}\label{sec:intro}

We report the results of searches for charged or neutral \B-meson decays to the 
charmless final states \cite{CC} \fetaKst,
\fetapKst, \fetarho, \fetaprho, \fetapiz, \fetappiz, \fomegapiz, and \fphipiz.
For decays that are self-tagging with respect to the $\b$ or $\bbar$ flavor,
we also measure the direct \CP-violating time-integrated charge asymmetry, 
\beqa
\acp =\frac{\Gamma^--\Gamma^+}{\Gamma^-+\Gamma^+}~.
\eeqa
The superscript on $\Gamma$ corresponds to the sign of the $B^\pm$ meson or the 
sign of the charged kaon for \Bz\ decays.
Throughout this paper, we use \etaprp\ to indicate either $\eta$ or \etapr.

\begin{figure}[!b]
\includegraphics[angle=0,width=\linewidth]{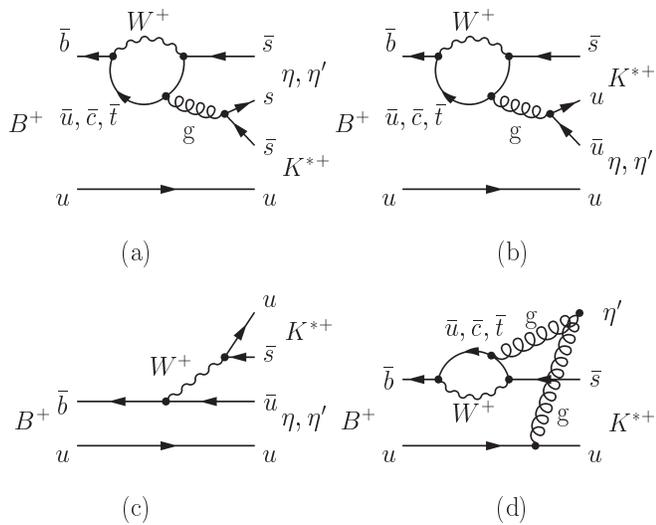}
\caption{\label{fig:feyn_etakstar} Feynman diagrams for the decays
$Bp\ra (\eta,\etapr)\Kstarp$.  The corresponding neutral decays are similar 
except that the spectator quark becomes a $d$, the gluon in (b) makes $d \bar d$,
and the tree diagram in (c) has an internal $W$.
}
\end{figure}

Interest in $B$ decays to $\eta$ or \etapr\ final states intensified in 1997
with the CLEO observation of the decay \etapK\ \cite{CLEOetapobs}.  It had
been pointed out by Lipkin six years earlier \cite{Lipkin} that
interference between two penguin diagrams (see Figs.~\ref{fig:feyn_etakstar}a 
and \ref{fig:feyn_etakstar}b) and the known $\eta/\etapr$
mixing angle conspire to greatly enhance \etapK\ and suppress \etaK.  
Because the vector \Kstar\ has the opposite parity from the kaon,
the situation is reversed for the \etapKst\ and \etaKst\ decays.  The
general features of this picture have already been verified by previous 
measurements and limits.  However, the details and possible contribution 
of the flavor-singlet diagram (Fig.~\ref{fig:feyn_etakstar}d) 
can only be tested with the measurement 
of the branching fractions of all four $(\eta,\etapr)(K,\Kstar)$ decays;
the branching fraction of the \etapKst\ decay is expected to be particularly 
sensitive to a flavor-singlet component \cite{chiang,beneke}.  
The tree diagram (Fig.~\ref{fig:feyn_etakstar}c) is suppressed by the
parameter $\lambda$ of the Cabibbo-Kobayashi-Maskawa (CKM) mixing matrix.

\begin{figure}[!b]
\includegraphics[angle=0,width=\linewidth]{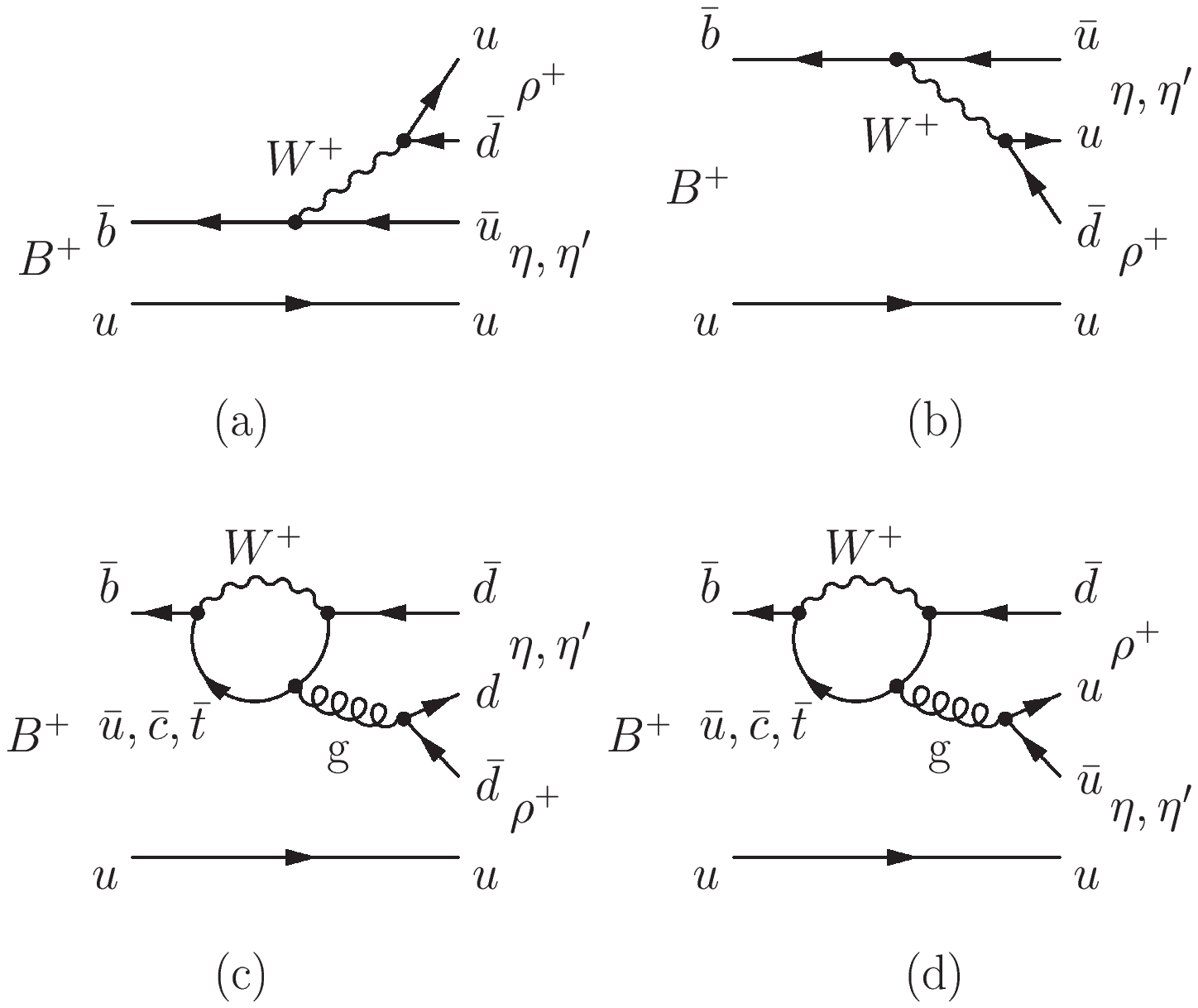}
\caption{\label{fig:feyn_etapirho} Feynman diagrams for the decays
\etarhop\ and \etaprhop.
}
\end{figure}

By contrast, for the $B\to\etaprp\rhop$ decays, the penguin diagrams 
(Figs.~\ref{fig:feyn_etapirho}c and \ref{fig:feyn_etapirho}d) are 
CKM- suppressed.  Since the internal tree diagram (Fig.~\ref{fig:feyn_etapirho}b)
is color-suppressed, the decay is dominated by the (external) tree diagram of
Fig.~\ref{fig:feyn_etapirho}a.

\begin{figure}[!hb]
\includegraphics[angle=0,width=\linewidth]{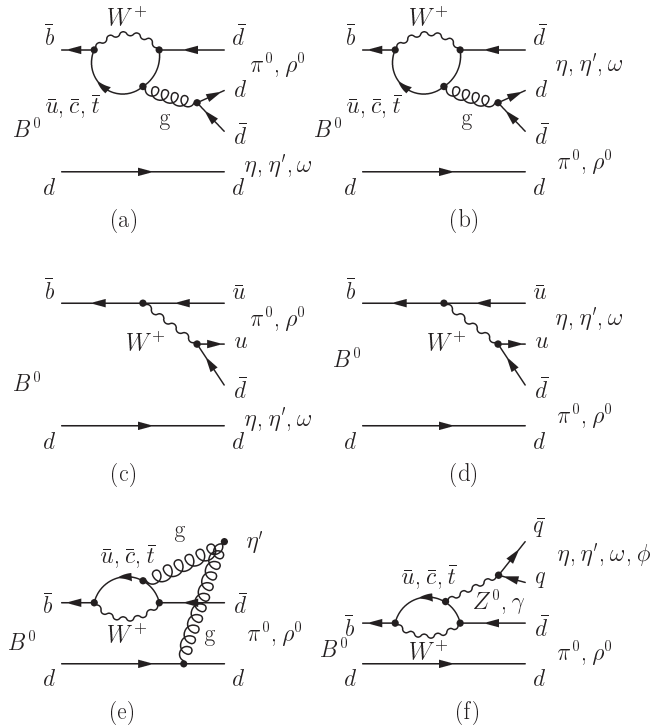}
\caption{\label{fig:feyn_pizrhoz} Feynman diagrams for the $\Bz$ decays.
}
\end{figure}

The \Bz\ decays are different because there are no external tree
diagrams analogous to Fig.~\ref{fig:feyn_etapirho}a.  In Figs.~\ref{fig:feyn_pizrhoz}a and
\ref{fig:feyn_pizrhoz}b we show the penguin diagrams and in 
Figs.~\ref{fig:feyn_pizrhoz}c and \ref{fig:feyn_pizrhoz}d the color-suppressed
tree diagrams for the $\Bz\ra\etaprp\rhoz$, $\Bz\ra\etaprp\piz$, and \omegapiz\ 
decays.  The color-suppressed diagrams cancel for the $\eta$ and \etapr\
decays and are expected to be largely suppressed for the pseudoscalar-vector 
($PV$) \omegapiz\ decay. The singlet penguin diagram (Fig.~\ref{fig:feyn_pizrhoz}e)
may be significant only for the decays with an \etapr\ in the final state, 
and the electroweak penguin 
(Fig.~\ref{fig:feyn_pizrhoz}f) is the only contribution for the \phipiz\ decay
(and negligible for the other decay modes).  Branching fractions for all 
these decays are generally expected to be in the range 
(0.1--10)$\times10^{-6}$ \cite{kramer,grabbag,yang,CGR}, with the 
$\Bp\ra\etaprp\rhop$ decays at the high end of this range and the \Bz\ decays 
at the low end (and \phipiz\ perhaps somewhat below this range).

The charge asymmetry \acp\ for most of these decays is expected to be
$\lsim10$\% \cite{kramer,AKLasym}.  However,
for \etapKst\ the penguin and tree amplitudes are expected 
to be of similar magnitude, which allows charge asymmetries which could
be in the $(20-40)$\% range
\cite{beneke,yang,CGR,dighe}.  Information 
on charge asymmetries and branching fractions from this full collection
of $B$ decays can serve to constrain the relationship between the various 
underlying amplitudes.

The results described in this paper complete the
measurement of all four $(\eta,\etapr)(K,\Kstar)$ final states, as well as
those with $(\eta,\etapr)(\pi,\rho)$, with a \babar\ dataset of 89
million \BB\ decays.  Current knowledge of the decays discussed here comes from 
published measurements from CLEO~\cite{CLEOetapr, CLEOomega, CLEOphipiz} and
\babar~\cite{BABARetapOmega}. Results for the final states
$(\eta,\etapr)(K,\pi)$ on this dataset have been presented
elsewhere~\cite{etaprPRL,etaomegaPRL}.  These data represent an order
of magnitude increase in the $B$ meson sample size over the only previous
complete study.

All results are based on extended maximum likelihood (ML) fits as 
described in Section \ref{sec:mlfit}.  In each
analysis, loose criteria are used to select events likely to contain
the desired signal $B$ decay.  A fit to kinematic
and topological discriminating variables is used to differentiate between
signal and background events and to determine signal event
yields and time-integrated rate asymmetries.  In all of the decays
analyzed, the background is dominated by random particle combinations
in continuum ($e^+e^-\to q{\bar q}$, $q=u,d,s,c$ ) events.  Some decay modes
also suffer backgrounds from other charmless $B$ decays with topologies
similar to that of the signal.  In such cases, these backgrounds are
accounted for explicitly in the fit as discussed in Sec. \ref{sec:bbbar}.
Signal event yields are
converted into branching fractions via selection efficiencies
determined from Monte Carlo simulations of the signal as well as
auxiliary studies of the data.  The complete analysis is carried out
without regard to whether there are observed signals.  This ``blind"
procedure is used to avoid bias in the results.

\section{Detector and Data} \label{sec:detector}

The results presented in this paper are based on data collected
with the \babar\ detector~\cite{BABARNIM}
at the PEP-II asymmetric-energy $e^+e^-$ collider~\cite{pep}
located at the Stanford Linear Accelerator Center.  The results in this
paper correspond to an accumulated integrated luminosity
of approximately 82~\invfb, corresponding to 89 million \BB\ pairs, 
recorded at the $\Upsilon (4S)$ resonance
(``on-peak'', center-of-mass energy $\sqrt{s}=10.58\ \gev$).
An additional 9.6~\invfb\ were recorded about 40~MeV below
this energy (``off-peak'') for the study of continuum backgrounds in
which a light or charm quark pair is produced.

The asymmetric beam configuration in the laboratory frame
provides a boost of $\beta\gamma = 0.56$ to the $\Upsilon(4S)$.  This results
in a charged-particle laboratory momentum spectrum from $B$ decays with an
endpoint near 4 \gev.
Charged particles are detected and their momenta measured by the
combination of a silicon vertex tracker (SVT), consisting of five layers
of double-sided detectors, and a 40-layer central drift chamber,
both operating in the 1.5-T magnetic field of a solenoid.  The transverse 
momentum resolution for the combined tracking system is 
$\sigma_{p_T}/p_T=0.0013p_T\oplus 0.0045$, where the sum is in quadrature and 
$p_T$ is measured in \gev.  For charged particles within the detector
acceptance resulting from the $B$ decays studied in this paper, 
the average detection efficiency is in excess of 96\% per particle.
Photons are detected and their energies measured by a CsI(Tl) electromagnetic
calorimeter (EMC).  The photon energy resolution is 
$\sigma_{E}/E = \left\{2.3 / E(\gev)^{1/4} \oplus 1.9 \right\}
\%$, and the angular resolution from the interaction point is
$\sigma_{\theta} = 3.9^{\rm o}/\sqrt{E(\gev)}$. The photon energy
scale is determined using symmetric $\piz\to\gamma\gamma$ decays.
The measured $\pi^0$ mass resolution for $\piz$'s with
laboratory momentum in excess of 1 \gev\ is approximately 8 MeV.

Charged-particle identification (PID) is provided by the average 
energy loss (\dedx) in the tracking devices and
by an internally reflecting ring-imaging 
Cherenkov detector (DIRC) covering the central region.  The $\dedx$ resolution 
from the drift chamber is typically about $7.5\%$ for pions.  The
Cherenkov angle resolution of the DIRC is measured to be 2.4~mrad, which
provides a nearly $3\sigma$ separation between charged kaons and pions at a
momentum of $3~\gev$.
Additional information
that we use to identify and reject electrons and muons is provided
by the EMC and the detectors
of the solenoid flux return (IFR).

\section{Candidate Reconstruction and \bma{B} Meson Selection}
\label{sec:eventsel}

We reconstruct $B$ mesons in the final states
\etaprp\Kstp, \etaprp\Kstz, \etaprp\rhop, \etaprp\rhoz, \etaprp\piz, 
\fomegapiz, and \fphipiz.  
Monte Carlo (MC) simulations \cite{geant}\ of the signal decay modes
and of continuum and \BB\ backgrounds, and data control samples of similar
modes, are used to establish the event
selection criteria.  The selection is designed to achieve high
efficiency and retain sufficient sidebands in the discriminating variables to
characterize the background for subsequent fitting.  As the invariant
mass distributions from the primary resonances (\etaprp , \Kst ,
$\rho$, \om , and $\phi$) in the decay are included in the maximum
likelihood fit, the selection criteria are generally loose.
Additional states---\piz\ or $\eta$ in \etapr\ decays, and
\kzs---are selected with the requirement that the invariant mass lie within
2-3$\sigma$ of the known mass.

\subsection{Charged track selection}
\label{sec:track_sele}

We require all charged-particle tracks (except for those from the
$\kzs\ra\pip\pim$ decay) used in reconstructing the $B$ candidate to 
include at least twelve point measurements in the drift chamber, lie 
in the polar angle range $0.41<\theta_{lab}<2.54$ rad, and 
originate from within 1.5 cm in the $x-y$ plane and 10 cm
in the $z$ direction from the nominal beam spot.  We require
the tracks to have a transverse momentum $p_T$ of at least 100 MeV.

\begin{figure}[!hb]
\includegraphics[angle=0,width=\linewidth]{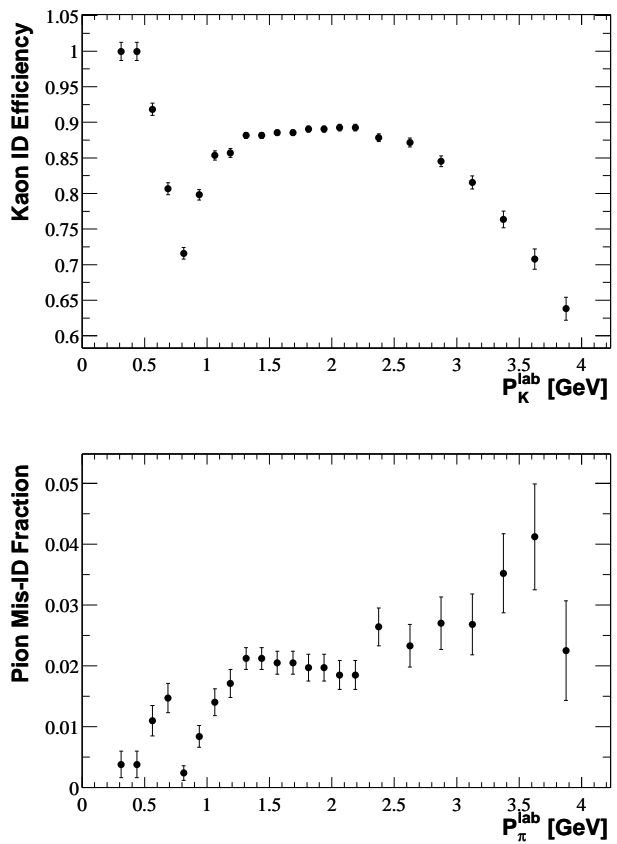}
\caption{\label{fig:kideff} Identification (ID) efficiency of the charged kaon 
selection as a function of the kaon laboratory momentum $P_{K}^{\rm lab}$ (top),
and fraction of charged pions misidentified (mis-ID) as kaons
as a function of the pion laboratory momentum $P_{\pi}^{\rm lab}$ (bottom).
The error bars
represent statistical uncertainties in the control sample of kaons and
pions from $D^*\to D^0\pi$, $D^0\to K\pi$ decays.}
\end{figure}

We also place requirements on particle identification criteria.  We veto
leptons from our samples by demanding that tracks have DIRC, EMC
and IFR signatures that are inconsistent with either electrons or muons.  
The remaining tracks are assigned as either charged pion or kaon candidates.  
This assignment is based on a likelihood selection developed from \dedx\ and 
Cherenkov angle information from the tracking detectors and DIRC, respectively.
For the typical laboratory momentum spectrum of the signal kaons, this  
selection has an efficiency of about 85\% and a pion misidentification rate of 
less than 2\%, as determined from control samples of $D^*\to D^0\pi$,
$D^0\to K\pi$ events.  The detailed performance of the kaon 
selection has been characterized as a function of laboratory momentum and 
can be seen in Fig.~\ref{fig:kideff}.

\subsection{\bma{\etaprp,~\om,~\rm{and}~\phi} selection}
\label{sec:etaom_sele}

We reconstruct the $\eta$ in two final states:
$\eta\ra\gamma\gamma$ ($\eta_{\gamma\gamma}$) and
$\eta\ra\pi^+\pi^-\pi^0$ (\etappp).  For the \etapr , we 
reconstruct two final states: \etaprrg\ (\etaprg ) and \etaptoepp\
(\etapepp ), with \etatogg\ (except in the \fetapfivepiKstz\ mode,
where we also include \etatoppp).  In the $\omegapiz$ channel, we
reconstruct $\omega\ra\thrpi$; for \phipiz\ we reconstruct
\phitoKpKm .  We place the following requirements on the invariant masses 
of the resonance candidates (in MeV): $520<m_{\eta_{3\pi}} < 570$,
$490<m_{\eta_{\gamma\gamma}}<600$, $910<m_{\etapr} <1000$ for 
\etaprg\ and \etapepp , $735<m_{\omega}<835$, and $990<m_{\phi}<1050$.
These ranges can be seen graphically in Fig.~\ref{fig:resmass} in
Sec.~\ref{sec:control}.  The mass requirements for these resonances are 
loose to keep appropriate sidebands for fitting; the resonance shapes
used for fitting are discussed in Sec.~\ref{sec:pdfparms}.

For \etatogg\ candidates we require $|\hel_\eta|$ to be less than 0.86, where
$\hel_\eta$ is the cosine of the $\eta$ decay angle. The decay angle is defined,
in the $\eta$ rest frame, as the angle between
one of the photons and the direction of the boost needed to get to this
frame from the $B$ center-of-mass (CM) frame.  This requirement removes 
very asymmetric decays of the $\eta$, where one photon carries most of the
particle's energy.   It is
effective against high-energy background photons from $B\ra K^*\gamma$
that combine with a random low-energy photon to form
an invariant mass in the range chosen for the \etatogg\ decay.
For the \fetapepprhop\ channel, the $\etatogg$ mass range is tightened
to $510< m_{\gaga} < 580$ \mev\ to reduce the continuum
background in the sample.

\subsection{Photon and \bma{\piz} selection}
\label{sec:gampiz_sele}

Photons are reconstructed from energy depositions in the electromagnetic
calorimeter which are not associated with a charged track.  We
require that all photon candidates have an energy greater than 30 \mev\ except
for the modes $\etaprp\piz$, \fomegapiz, and \fphipiz, where there is 
significant combinatorial background arising from low-energy photons.  For
these modes, we tighten the photon-energy requirement to 50 \mev\ for all 
photons. For \etatogg, we require each photon energy to be greater than 
100 \mev, and for the \etaptorg\ modes, we 
require the photon from the \etapr\ decay to exceed 200 \mev.

We select neutral-pion candidates from two photon clusters with the 
requirement that the \gaga\
invariant mass satisfy $120 < m_{\pi^0} < 150$ \mev.  The mass of a \piz\
candidate meeting this criterion is then constrained to the nominal
value~\cite{PDG2002} and, when
combined with other tracks or neutrals to form a $B$ candidate, to
originate from the $B$ candidate vertex.  This 
procedure improves the mass and energy resolution of the parent particle.

For the primary \piz\ in $\etaprp\piz$ decays, 
photon candidates are required to be consistent with the expected
lateral shower shape, and the magnitude of the cosine of the \piz\ 
decay angle (defined as for the $\eta$) must be less than 0.95.

\subsection{\bma{\Kz} selection}
\label{sec:kz_sele}

For decay chains containing a \Kz, we reconstruct only the
$\kzs\ra\pip\pim$ decay.  The invariant mass of the candidate 
\kzs\ is required
to lie within the range $488 < m_{\pi^+\pi^-} < 508$ \mev.  We also perform a
vertex-constrained fit to require that the two
tracks originate from a common vertex, and require that the lifetime
significance of the \kzs\ ($\tau/\sigma_{\tau}$) be $>3$, where
$\sigma_{\tau}$ is the uncertainty in the lifetime determined from the
vertex-constrained fit.

\subsection{\bma{\Kst} and \bma{\rho} selection}
\label{sec:kstrho_sele}

We reconstruct the \Kstarp\ as either $K^+\piz$ (\KstpKppiz ) or 
$\KS\pip$ (\KstpKspip ), and the \Kstz\ as $K^+\pim$ (\KstzKppim ).
The $\rho^+$ is reconstructed as $\pip\piz$ and the
$\rho^0$ as $\pip\pim$.  A vertex fit is performed when reconstructing the 
resonant \Kstar\ or $\rho$ candidate.  We require the invariant
masses (in MeV) of the resonance candidates to be in the ranges:
$755<m_{K\pi}<1035$, $470<m_{\pi^+\piz}<1070$, and $510<m_{\pi^+\pi^-}<1060$.
The lower limit on the \rhoz\ candidate invariant mass is chosen to reject
background from \kzs\ decays.  

For decay chains involving a charged \Kst\ or $\rho$,
we define \hel, the cosine of the angle between the pion 
and the negative of the $B$ momentum in the vector-meson rest frame.
For \rhop\ decays, the direction is that of the \piz.  For \rhoz\
decays, we use only the magnitude of \hel, which is independent of the choice
of reference pion.
For these decays with a \piz\ in the final state, we require that \hel\ 
be greater than $-0.5$ to reject combinatorial background.

\subsection{\bma{B} meson selection}
\label{sec:bmeson_sele}

A $B$-meson candidate is characterized kinematically by the
energy-substituted mass \mes\ and by the energy difference \DE,
defined as 
\begin{eqnarray}
\mes &=& \sqrt{\left(\frac{\half s + \pvec_0\cdot \pvec_B}{E_0}\right)^2 - \pvec_B^2} \qquad \mbox{and}  \\
\DE  &=& \left(2q_0q_B-s\right)/2\sqrt{s} \ ,
\end{eqnarray}
where $q_B=(E_B,\pvec_B)$ and $q_0=(E_0,\pvec_0)$ are the four vectors 
of the $B$-candidate and the initial electron-positron system, respectively, 
and $s$ is the square of the invariant mass of the electron-positron system.  
When expressed in the
\UfourS\ frame, these quantities take the simpler but equivalent form
\begin{eqnarray}
\mes &=& \sqrt{{1\over4}s - \pvec_B^{*2}} \qquad \mbox{and}  \\
\DE  &=& E_B^*-\half\sqrt{s} \ ,
\end{eqnarray}
where the asterisk denotes the value in the \UfourS\
frame. The mode-dependent resolutions on these quantities for
signal events are about $3\ \mev$ for \mes , and 30--60 MeV for \DE.

We require $5.20\le\mes\le5.29\ \gev$ and $|\DE|\le0.2$ \gev\ for all but
the $\etaprp\piz$, \fomegapiz , and \fphipiz\ modes, where we loosen
the \DE\ range to $|\DE|\le0.3$ \gev\ to account for poorer detector
resolution in these channels.

When multiple $B$ candidates from the same
event pass the selection requirements, we choose a single candidate
based on criteria described below.  The average number of candidates per event 
depends on the mode; it is typically about 1.2 and is always less than 1.5.  
We find that 70--90\% of the events have a single combination and about
90\% of the rest have two combinations.  In decays containing an 
$\eta$  and a \Kst\ or $\rho$, we select the candidate with the smallest 
$\chi^2$ formed from the $\eta$ and \Kst\ or $\rho$ masses. For decays 
containing \etaptoepp, the $\chi^2$ is formed from the masses of the 
\etapr\ and $\eta$ candidates.
For all other decays, we retain the candidate that has the mass of the 
primary resonance (\etaprp , \om , or $\phi$) closest to the nominal value
\cite{PDG2002}.  We have checked that this choice introduces no significant
yield bias, in part because, for the primary resonance mass, there is an
adjustable peaking component included in the fit, which would account for 
any small distortion due to this selection.

\section{Sources of Background and Suppression Techniques} \label{sec:bkg}

Production of $B{\bar B}$ pairs accounts for a relatively small fraction 
of the \epem\ cross section even at the peak of the \UfourS\ resonance.
Upsilon production amounts to about 25\% of the total hadronic cross 
section, while tau-pair production and other QED processes occur as well.  
We describe below several sources of background, and discuss techniques for 
distinguishing them from signal.

\subsection{QED and tau-pair backgrounds}

Two-photon processes, Bhabha scattering, muon-pair production and tau
pair production are characterized by
low charged track multiplicities.  Bhabha and muon-pair events
are significantly prescaled at the trigger level.  We further
suppress these and other tau and QED processes via a minimum
requirement on the event track multiplicity.  We require the event to
contain at least one track more than the topology of our final state,
or three tracks, whichever is larger.  We also place a requirement on the 
ratio of the second to the zeroth Fox--Wolfram moments \cite{FoxW}, $R_2<0.98$,  
calculated with both charged tracks and neutral energy depositions.  These 
selection criteria are more than 90\%  efficient when applied to signal.
From MC simulations we have determined that the remaining background
from these sources is negligible.

\subsection{QCD continuum backgrounds}

The primary source of background to all charmless hadronic decays of the 
$B$ meson arises from continuum quark-antiquark production.  The fact that 
these events are produced well above threshold provides the means by which
they can be rejected, as the hadronization products are produced in a jet-like 
topology.  In strong 
contrast, $B$ mesons resulting from \UfourS\ decays are produced just above
threshold.  Thus the final-state particles in the signal 
are distributed approximately isotropically in the CM frame.  

\begin{figure}[!hb]
\includegraphics[angle=0,width=\linewidth]{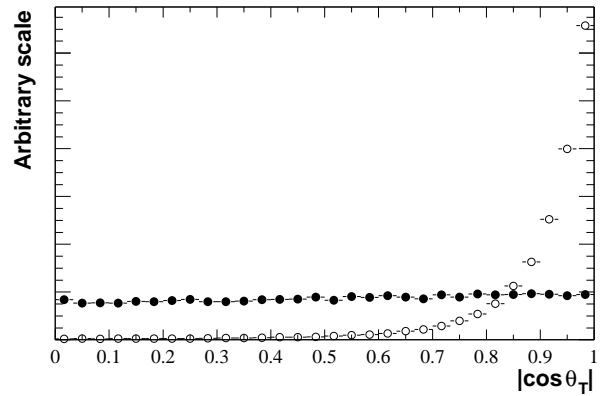}
\caption{\label{fig:thrust} Distribution in $|\costhr|$ for a typical
$B$ meson decay (\etaprgpiz\ MC, solid points) and for the corresponding 
continuum background data (open circles).
}
\end{figure}

Several event-shape variables are designed to take advantage of this
difference.  We define the thrust axis for a collection of particles as the axis
that maximizes the sum of the magnitudes of the longitudinal momenta with 
respect to the axis.  The angle $\theta_T$ 
between the thrust axis of the $B$ candidate and that of
the rest of the tracks and neutral clusters in the event, calculated in
the \UfourS\ frame, is the most powerful of the shape variables we employ.  The 
distribution of the magnitude of \costhr\ is
sharply peaked near 1 for combinations drawn from jet-like \qqbar\
pairs and is nearly uniform for the isotropic $B$-meson decays.  This behavior
is shown in Fig.~\ref{fig:thrust}.  The selection
criterion placed on \costhr\ is optimized for each channel to maximize 
our sensitivity to signal in the presence of continuum background and to reduce 
the size of the sample entering the fit.  The optimization procedure is 
described in Sec.~\ref{sec:validation}.  The maximum allowed value of $|\costhr|$
chosen for each signal mode is listed in Table~\ref{tab:sumtab_all}.

\begin{figure}[!htb]
\includegraphics[angle=-90,width=\linewidth]{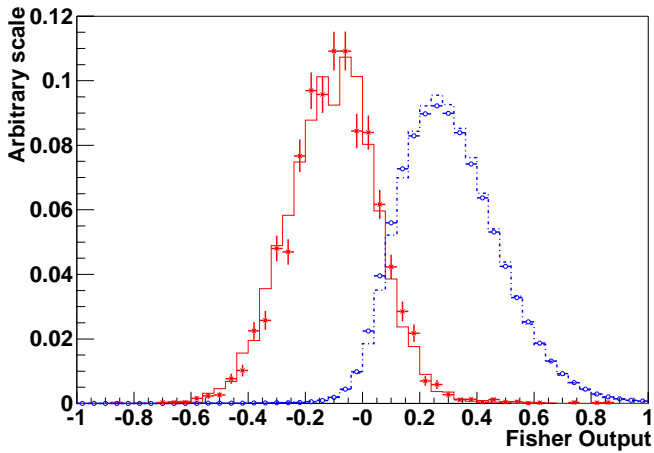}
\caption{\label{fig:fisher} Distributions of Fisher-discriminant output for the
data control mode $B^-\ra D^0\pi^-,\;D^0\ra K^-\pi^+\pi^0$ (points with error 
bars), corresponding signal Monte Carlo (solid
histogram), continuum data (open circles) and continuum Monte Carlo 
(dashed histogram) after requiring
$|\costhr|<0.9$.  The Fisher discriminant and $|\costhr|$ are strongly
correlated, so the separation depends on this requirement.
}
\end{figure}

Further use of the event topology is made via the construction of a Fisher 
discriminant \xf, which is subsequently used as a discriminating variable in
the likelihood fit.  The Fisher discriminant we use is an optimized
linear combination of the remaining event shape information (excluding
\costhr, which we have already used in our preselection
requirements).  The variables entering the Fisher discriminant are the
angles with 
respect to the beam axis of the $B$ momentum and $B$ thrust axis (in
the \UfourS\ frame), and the zeroth and second angular moments
$L_{0,2}$ of the energy flow about the $B$ thrust axis.  The moments
are defined by $ L_j = \sum_i p_i\times\left|\cos\theta_i\right|^j,$
where $\theta_i$ is the angle with respect to the $B$ thrust axis of
track or neutral cluster $i$, $p_i$ is its momentum, and the sum
excludes the $B$ candidate.  The coefficients used to combine these
variables are chosen to maximize the separation (difference of means
divided by quadrature sum of errors) between the signal and continuum 
background distributions of $ L_j$, and are determined from studies of signal 
MC and off-peak data.  We have studied the optimization of \xf\ for a variety
of signal modes, and find that the optimal sets of coefficients are
nearly identical for all.  Thus we do not re-optimize the Fisher
coefficients for 
each individual decay.  Because the information contained in \xf\ is
correlated with $|\costhr|$, the separation 
between signal and background is dependent on the $|\costhr|$
requirement made prior to 
the formation of \xf.  In Fig.~\ref{fig:fisher}, we show the Fisher-discriminant
distribution for signal and continuum background for the $B^-\ra D^0\pi^-$ 
control sample.

\subsection{\bma{\BB} backgrounds}
\label{sec:bbbar}

Most charmless hadronic-$B$-decay analyses do not have much background from 
other $B$ decays.  Specifically, since most $B$ mesons decay via $b\to c$ 
transitions, the strange and light meson decay products from such decays result
from $b\to c\to q$ cascades, and thus have lower momentum than those expected in
the signal final states.  This small background is included in our
\qqbar\ background PDF shapes (see next section) since the shapes are
extracted from on-peak data.

We have found, however, that some of the signal modes (see Table 
\ref{tab:restab_all} in Sec.~\ref{sec:results}) do suffer from
backgrounds from charmless hadronic decay modes.  We investigate backgrounds 
that may not be completely suppressed by the selection criteria defined in 
Sec.~\ref{sec:eventsel} with Monte Carlo samples of \BB\ events
corresponding to several times the number of such events in the dataset.  
When we find an indication of a high selection rate for a 
particular background decay mode, we use the experimentally measured 
(when available) or theoretically predicted branching fraction of that
mode to determine its expected contribution.  Fits to simulated
experiments such as those described in Sec.~\ref{sec:validation} are 
used to evaluate whether such events cause a significant bias
to the measured signal yield.  Based on these studies, we have adjusted 
(while still blind) some selection criteria and in some cases added a
component to the ML fit to account explicitly for the remaining \BB\
background contributions. Systematic errors account for the uncertainties in
this method.  The details of this procedure are described below. 

\section{Maximum Likelihood Fit}\label{sec:mlfit}

We use an unbinned, extended maximum likelihood fit to extract
signal yields for our modes.  A subsample of events to fit for each
decay channel is selected as described in Sec.~\ref{sec:eventsel}.
The sample sizes for the decay chains reported here range from 700 to
30,000 events, where we include sidebands in
all discriminating variables in order to parameterize the backgrounds.

\subsection{Likelihood function} \label{sec:like}

The likelihood function incorporates several discriminating variables to
distinguish signal from the large number of background events retained
by the sample selection.  We describe the $B$-decay kinematics with
two variables: \DE\ and \mes\ (as defined in
Sec.~\ref{sec:bmeson_sele}).  We also include the mass of the primary 
resonance candidate ($m_{\eta}$, $m_{\etapr}$, $m_{K^*}$, $m_\rho$, $m_{\om}$, 
or $m_{\phi}$) and the Fisher discriminant \xf.
For the vector-pseudoscalar modes with a $K^*$, $\rho$, \om, or $\phi$, 
we also include in the fit the helicity cosine \hel\ of the vector meson.
For the $\Kstar$, $\rho$, and $\phi$, \hel\ is defined in
Sec.~\ref{sec:kstrho_sele}.
For the \omegapiz\ decay, \hel\ is defined as the cosine of the 
angle between the normal to the $\omega$ decay plane
(the plane of the three pions in the $\omega$ rest frame) and the
flight direction of the $\omega$, measured in the $\omega$ rest frame.

Because correlations among the discriminating variables (except resonance mass
and \hel\ for background) in the selected data are 
small, we take the probability distribution function (PDF) for 
each event $i$ to be a product of the PDFs for the separate
discriminating variables.  
We define hypotheses $j$, where $j$ can be signal, continuum
background, or (where appropriate) \BB\ background. The PDFs can be
written as  
\begin{equation}
{\cal P}^i_{j} =  
{\cal P}_j (m_{ES}^i)  {\cal  P}_j (\DE^i)  { \cal P}_j(\xf^i) 
{\cal P}_j (m^i_P)  {\cal P}_j (m^i_V,{\cal H}^i),
\end{equation}
where $m_P$ indicates the pseudoscalar candidate mass in the fit (absent for
\omegapiz\ and \phipiz\ modes) and $m_V$
indicates the vector candidate mass (absent for the $\etaprp\piz$ modes).

The likelihood function for each decay mode is

\begin{equation}
{\cal L} = \frac{\exp{(-\sum_j Y_{j})}}{N!}\prod_i^{N}\sum_j Y_{j} {\cal P}^i_{j}\,,
\end{equation}

\noindent where $Y_{j}$ is the yield of events for hypothesis $j$ (to be
found by the fitter) and $N$ is the observed number of events in the sample.  The
first factor takes into account the Poisson fluctuations in the total
number of events.

\section{Signal and background model}
\label{sec:pdfparms}

We determine the PDFs for signal from MC distributions for each
discriminating variable.  The PDFs for \BB\ background (where appropriate) arise
from fitting the composite \BB\ MC sample, described in
Sec.~\ref{sec:bbbar_pdfs}.  For the continuum background we
establish the functional forms and initial parameter values of the
PDFs with data from sidebands in \mes\ or \DE.  We then refine the
main background parameters (excluding resonance-mass central values
and widths) by allowing them to float in the final fit so that they are
determined by the full data sample.  
The following sections describe first the construction of samples to represent 
\BB\ background, and then the control samples used to validate the PDF shapes 
and make adjustments to the means and widths of the distributions where needed.
Finally we describe the detailed functional forms
used to parameterize all of the signal and background distributions.

\subsection{Inclusion of \bma{\BB} background in the fits}
\label{sec:bbbar_pdfs}

As discussed in Sec.~\ref{sec:bbbar}, backgrounds from other
charmless $B$ decays need to be accounted for explicitly in the
maximum likelihood fit for some decay chains.  

Since we find that the signal yield bias due to \BB\ background for the \etappp\Kst\
channels is less than 1\% of the signal yield, we do not
include a \BB\ component for these modes.  For all modes with a
$\Kstarp\to\Kp\piz$ decay, nearly all \BB\ backgrounds are removed by
the requirement $\calH>-0.5$.  This requirement is also helpful in 
reducing the \BB\ background for decays with a $\rhop\to\pip\piz$, 
though sufficient background remains to be included in the fit.  For all 
other modes except \phipiz, we include a \BB\ component in the fit.
The fit number of \BB\ events is a small fraction of the total sample 
and is tabulated in Table \ref{tab:restab_all} in Sec.~\ref{sec:results}.

The PDFs for \BB\ background are determined by fitting a 
sample of MC events composed of several charmless decay
chains, with the PDF shapes described below.
For the \etatogg\ channels, the \BB\ background is dominated by
$\B\ra\Kst\gamma$ decays, even after the $\eta$ decay angle requirement, 
due to the relatively large $K^*\gamma$ branching fraction ($40\times10^{-6}$).
For the \etarho\ channels, the largest backgrounds are from \fetaKst\ decays,
with misidentification of the charged kaon or loss of the kaon while
selecting a pion from the other $B$.  For the \etapr\ channels, the 
dominant backgrounds in all modes, except for \fetaprgrho, arise from
\etapK\ decays, due to the relatively large branching fraction 
($\sim70\times 10^{-6}$).  Another important background for the \etaprg\Kst\ 
channels, is $\Kst\rhoz$ decays, where the $\rho$ is combined with a photon 
to fake an $\etapr$.  For the \fetaprgrho\ and \fetaprgpiz\ modes, \BB\ 
backgrounds are primarily from $\Bp\to\rhop\rhoz$ and $\Bz\to\rhop\rhom$ decays.
For the decays with a primary \piz, the largest backgrounds are from 
$\Bp\to\etaprp\rhop$ and $\Bp\to\omega\rhop$ decays, where due to the 
forward-backward peaking of the \rhop\ \hel\ distribution, 
the \piz\ is often energetic and the charged pion is lost.

\subsection{PDF corrections from data control samples}
\label{sec:control}

\begin{figure}[htb!]
\vspace{0.5cm}
 \includegraphics[angle=0,width=\linewidth]{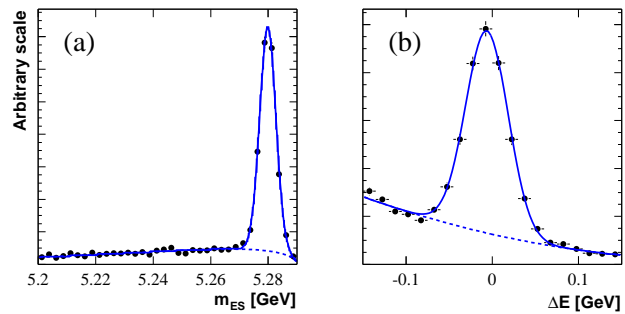}
 \caption{\label{fig:mesde} Distributions of (a) \mes\ and (b) \DE\
from the $\Bm\ra\pim\Dz$ data sample used to determine the small corrections to
signal Monte Carlo PDF shapes.}
\end{figure}

We validate the simulation on which we rely for signal PDFs by comparing critical
distributions of discriminating variables in MC with those from large data control samples.  
For \mes\ and \DE\ (see Fig.~\ref{fig:mesde}), we use
the decays \Dcontrol, which have similar topology to the modes under study 
here.  We select these samples by making loose requirements on \mes and \DE, and 
more stringent selections on \costhr\ and the $D^0$ and $\rho$ candidate 
masses (as appropriate). 
We also place kinematic requirements on the $D$ and $B$ daughters to force the
charmed decay to look as much like that of a charmless decay as 
possible without eliminating the control-sample signal.  These selection 
criteria are applied both to the data and to a MC mixture of related
$B\ra DX$ and  $B\ra D^*X$ decays, which simulates the crossfeed from 
$D^*\ra D^0$ decays observed in data.  From these control samples, we
determine small adjustments
to the mean value of the signal \mes\ distribution and to the resolution of 
the \DE\ distribution compared with Monte Carlo.  For \xf\ we use
parameters found from a sample of approximately 500 \etaprgKp\ events,
with a \costhr\ requirement matching that used for each signal mode.

\begin{figure}[htb!]
\vspace{0.5cm}
 \includegraphics[angle=0,width=\linewidth]{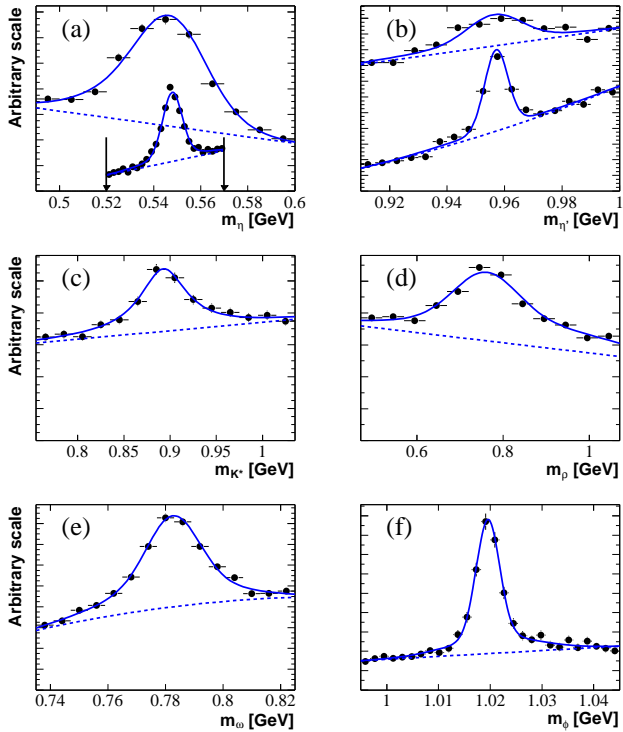}
 \caption{\label{fig:resmass} Distributions of the candidate masses for
resonant decays from the on-peak sideband samples in data that are used to 
describe the signal PDF shapes (see Secs. \ref{sec:mPpdf} and
\ref{sec:mV_helpdf}).  For each distribution a real resonance signal 
component is evident above a combinatorial background component: 
(a and b) the four \etaprp\ candidate mass combinations from the
\fetarho\ and \fetaprho\ samples;
(c) \Kst\ candidate mass from the \fetaprgKst\ sample;
(d) primary $\rho$ candidate mass from the \fetaprgrho\ sample;
(e) \om\ candidate mass from the \fomegapiz\ sample; 
(f) $\phi$ candidate mass from the \fphipiz\ sample. 
In (a) the arrows indicate the narrower mass requirement for the
\etatoppp\ decay.  The same range is used even for the narrower 
\etaptoepp\ distribution, shown as the lower plots in (b).
For the \Kst\ and $\rho$ cases, we do not show both
charges since the distributions are very similar.}
\end{figure}

For the mass shapes of the resonances, we study inclusive resonance production 
in the off-peak data and corresponding continuum MC.  In each sample, we 
reconstruct resonance candidates involved in our final states, requiring a 
minimum value of the candidate CM momentum of 1.9 \gev\ to reflect the 
kinematics of our final
states.  The resolutions and means of the invariant mass distributions are
compared, and we adjust the means and widths of PDF parameterizations based 
on the outcome of these results.  A typical mass distribution for each resonance
is shown in Fig.~\ref{fig:resmass}.

\subsection{\bma{\mes} parameterization}
\label{sec:mespdf}

The signal distribution in \mes\ is parameterized by two Gaussian
functions centered near the mass of the $B$ meson.  The second Gaussian
typically accounts for less than 20\% of the total area, and
has a larger width to take into account the tails of the distribution,
which arise primarily from misreconstructed signal events.  In
continuum background, we model \mes\ by a phase-space-motivated
empirical function \cite{argus} of the form
\beqa
\label{eqn:argus}
f(x) \propto x\sqrt{1-x^2}\exp{\left[-\xi(1-x^2)\right]} \, ,
\eeqa
where we define $x\equiv 2\mes/\sqrt{s}$, and $\xi$ is a parameter determined 
by the fit.  In \BB\ background samples, we find that the \mes\ distribution 
is well-described by adding a simple Gaussian function to the empirical shape 
in Eq.~\ref{eqn:argus}; a
similar alternate form of a Gaussian convolved with an exponential is 
used for some channels.

\subsection{\bma{\DE} parameterization}
\label{sec:depdf}

For \DE, we fit the signal distribution with two Gaussian functions, both centered 
near zero.  The broad Gaussian has a width about five times larger than the 
narrow Gaussian; this accounts for energy loss before or leakage out of the EMC,
as well as incorrect candidate combinations in true signal events.  The 
broad Gaussian component becomes larger as more of the final state energy is
carried by neutral particles.  The primary Gaussian function
accounts for about (60--80)\% of the total area in all modes except 
$\etaprp\rho$ where it is  between 30\% and 60\%.  For continuum
background, we model the \DE\ distribution with a linear or quadratic
polynomial as required by the data.  The \BB\ background is
described well by two Gaussian functions peaking at negative (positive) \DE, 
accounting for backgrounds that have a larger (smaller) number of
tracks and neutrals in the final state than the signal.

\subsection{Fisher parameterization}
\label{sec:xfpdf}

For both signal and background, the Fisher distribution \xf\ is
described well by a Gaussian function with different widths to the
left and right of the mean.  For the continuum background
distribution, we also include a second Gaussian function with a larger
width to account for a small tail in the signal \xf\ region.  This additional
component of the PDF is important, because it prevents the background 
probability from becoming infinitesimally small in the region where
signal lies.  As shown in Fig.~\ref{fig:fisher}, the mean of
the continuum background distribution is approximately
$2\sigma$ greater than the mean of the signal peak, allowing for strong
discrimination between the two.  Because \xf\ describes the overall
shape of the event, the distribution for \BB\ background is
very similar to the signal distribution; hence this variable has
little discriminating power against \BB\ background.

\subsection{Pseudoscalar mass parameterization}
\label{sec:mPpdf}

The pseudoscalar candidate mass distributions for signal are described well by 
the sum of two Gaussian functions.  We use MC values for the means
and widths of these Gaussians, corrected where necessary by using samples such as
those shown in Fig.~\ref{fig:resmass}.  In continuum background, 
we fit the data with two Gaussian functions, where we fix
the means and widths to those used for signal, and include a linear or
quadratic term to account for non-resonant background.  The fraction
of resonant to non-resonant background is allowed to float in the
final fit.  When there is no discernible resonant component, as in
\etaptorg, floating this parameter can cause convergence issues in
the final ML fit.  If validation studies show this effect, the
resonant fraction is fixed in the final analysis.  For
\BB\ background, we use the same functional form as in continuum
background; whether or not there is a true resonant component in \BB\
background depends upon the charmless decay chains expected to
contribute.

\subsection{Vector mass and helicity parameterization}
\label{sec:mV_helpdf}

In pseudoscalar--vector decays of the $B$ meson, the vector meson has a 
helicity-angle distribution proportional to $\hel^2$ for true signal events.
We model the vector-meson helicity distribution for signal with a polynomial
times a threshold function that allows for the effects of acceptance.
The signal \Kst\ and \om\ invariant-mass distributions are described by 
Breit-Wigner shapes.  The $\phi$ and $\rho$ line 
shapes are found to be modeled well by two Gaussian functions; these do not 
fit well to a Breit-Wigner shape because of non-negligible mass resolution 
($\phi$) or misreconstructed $\rho$ candidates in real signal events ($\rho$).
For the $\rho$ and other wide distributions there is as much as 10\% loss of 
efficiency due to the effect of the mass range requirements; this effect
is included in the overall efficiency estimate and its uncertainty is
included in systematic errors discussed in Sec.~\ref{sec:syst}.
See Fig.~\ref{fig:resmass} for illustrations of these distributions.

Because the shape of the helicity angle can be different for continuum 
background with and without a true vector resonance, we use a 
two-dimensional PDF to describe the resonance mass distribution and the 
helicity-angle distribution.
We would expect that the background \hel\ would have a nearly uniform 
distribution, corresponding to a sum of combinatorial resonance background and
background of true resonances from various production mechanisms.  We find that 
the pure-background shape is modeled well by a second order polynomial with
only a small amount of curvature and the true-resonance component is a
separate low-order-polynomial shape.  The mass parameters for the 
true-resonance component are fixed to be the same as for the signal.

The \BB\ background component of \hel\ is modeled by a single fourth-degree 
polynomial.  We parameterize the resonance mass
distribution with two Gaussian functions plus a linear or quadratic
polynomial, allowing the means and widths of the Gaussians to float if
the resonant component of the background differs from the signal
resonance.  This is especially necessary when \BB\ background arises
when a misidentified kaon from a \Kst\ causes its reconstruction
as a $\rho$.  

The requirement that charged tracks have $p_T> 100$ \mev\
(Sec.~\ref{sec:eventsel}) can induce a ``roll-off"
effect near \hel\ values of $\pm 1$. In particular, for decays of a
\Kstar\ or $\rho$ with a charged pion, the helicity distribution of
the vector meson shows a characteristic roll-off in the region
populated by low-momentum pions.  This effect is absent for charged kaons
since there are no kaons with $p_T< 100$ \mev. We model the
roll-off in both the signal and background \hel\ distributions by
multiplying the primary PDF shape by an appropriate Fermi-Dirac
threshold function.  The parameters of this roll-off function are
constrained to be the same for signal and both background components.
Because the \om\ helicity angle is defined from a
three-body decay (\omtoppp ), there is little correlation between
low-momentum pions and helicity angle, and hence no significant roll-off.

\section{Fit Validation}
\label{sec:validation}

Before applying the fitting procedure to the data to extract the signal
yields we subject it to several tests.  Internal consistency is checked
with fits to ensembles of ``experiments" generated by Monte Carlo from
the PDFs.  From these we establish the number of parameters associated
with the PDF shapes that can be left free in addition to the
yields.  Ensemble distributions of the fitted parameters verify that 
the generated values are reproduced with the expected resolution.  The
ensemble distribution of $\ln{\calL}$ itself provides a reference to
check the goodness of fit of the final measurement once it has been
performed. 

We account for possible biases due to neglecting correlations among 
discriminating variables in the PDFs by fitting ensembles of experiments 
into which we have embedded the expected
number of signal events randomly extracted from the detailed MC
samples, where correlations are modeled fully.  We find 
a positive bias of a few events for most modes, as shown in
Table~\ref{tab:sumtab_all}.  Events from a weighted mixture of simulated \BB\
background decays are included where significant, and so the bias we
measure includes the effect of crossfeed from these modes.

\begin{table}[!tp]
\caption{For each \B\ decay chain we present the optimized $|\costhr|$
requirement, the number of on-peak events passing the preselection
requirements, and the fit bias $Y_b$ determined from simulated experiments
(the uncertainty on this bias is discussed in Sec.~\ref{sec:syst}).
}
\label{tab:sumtab_all}
\begin{center}
\vspace*{-0.3cm}
\begin{tabular}{lcrc}
\dbline
Mode          		&~Max $|\costhr|$~&~\#Events~& Fit bias, $Y_b$ \\
			&		& in fit~~~& (events)\\
\sgline
\etaKstp		& 		&	  &	\\
~~~\fetaggKstpKspip	& 	0.90	& 7573~~~~&\msp4.7\\
~~~\fetapppKstpKspip	& 	0.90	& 4132~~~~&\msp1.7\\
~~~\fetaggKstpKppiz	& 	0.90	& 4974~~~~&\msp0.1\\
~~~\fetapppKstpKppiz	& 	0.90	& 2835~~~~&\msp0.3\\
\etaKstz		&		&	  &	\\
~~~\fetaggKstz		& 	0.90	&12179~~~~&\msp8.1\\
~~~\fetapppKstz		& 	0.90	& 6440~~~~&\msp1.8\\
\etarhop		& 		&		&	\\
~~~\fetaggrhop		& 	0.80	&17084~~~~&\msp1.3\\
~~~\fetappprhop		& 	0.90	&16106~~~~&\msp1.0\\
\etarhoz		& 		&		&	\\
~~~\fetaggrhoz		& 	0.70	&11107~~~~&$-1.0$\\
~~~\fetappprhoz		& 	0.80	& 8347~~~~&\msp2.3\\
\etapiz			& 		&  	  &	\\
~~~\fetaggpiz		& 	0.80	& 5379~~~~&$-1.1$\\
~~~\fetappppiz		& 	0.80	& 2271~~~~&\msp0.7\\
\sgline
\etapKstp		& 		&		&	\\
~~~\fetapeppKstpKspip	& 	0.90	& 2973~~~~&$-4.5$\\
~~~\fetaprgKstpKspip	& 	0.75	&13299~~~~&\msp3.6\\
~~~\fetapeppKstpKppiz	& 	0.90	& 2009~~~~&\msp0.0\\
~~~\fetaprgKstpKppiz	& 	0.75	& 8205~~~~&\msp0.6\\
\etapKstz		& 		&	  &	\\
~~~\fetapeppggKstz	& 	0.90	& 4808~~~~&$-3.7$\\
~~~\fetapfivepiKstz	& 	0.90	&  695~~~~&\msp1.7\\
~~~\fetaprgKstz		& 	0.75	&20504~~~~&\msp4.2\\
\etaprhop		& 		&	&	\\
~~~\fetapepprhop	& 	0.90	& 8737~~~~&\msp2.1\\
~~~\fetaprgrhop		& 	0.65	&28933~~~~&\msp7.8\\
\etaprhoz		& 	0.90	& 9515~~~~&$-3.7$\\
\etappiz		& 		&		&	\\
~~~\fetapepppiz		& 	0.90	& 3491~~~~&$-3.5$\\
~~~\fetaprgpiz		& 	0.70    &11426~~~~&\msp2.8\\
\sgline
\omegapiz		& 	0.80    &18986~~~~&$-2.1$\\
\sgline
\phipiz			& 	0.90   	& 4840~~~~&$-1.1$\\
\dbline
\end{tabular}
\end{center}
\end{table}

For modes with low background and small signal yields, the ensemble
yield distribution may exhibit a significant negative tail. This is due
to the nature of the maximum likelihood method, which is known to be
biased for small samples. 
The source of the bias is the insufficient number of events for
which the probability for the signal hypothesis is larger than the probability
for the background hypothesis.
This results in a negative bias, which is taken as the mean of the yield
distribution from the fits to the ensembles described above. Examples of
modes with negative bias can be found in Table~\ref{tab:sumtab_all}.
By subtracting the bias we correct for this effect on average,
and we include the uncertainty as a systematic error.

This same procedure for generating and fitting simplified MC samples is
used to find an optimal selection requirement for the \costhr\ variable in
the early stages of each analysis.  The studies are performed for a range 
of selection values, to minimize the fractional error on the signal yield.
The optimal values of the \costhr\ requirement that are chosen are given in 
Table \ref{tab:sumtab_all}.

Finally, we apply the fit to the off-peak data to confirm that we
find no fake signals in a sample with no signal events.

\section{Efficiencies and Efficiency Corrections}
\label{sec:eff}

The efficiency is determined by the ratio of the number of signal Monte Carlo
events passing preselection to the total number of generated MC signal
events.  This efficiency is corrected for differences between the
true detector efficiencies and those simulated in Monte Carlo.  From a
study of absolute tracking efficiency, we apply a correction of (1--7)\%,
depending on the number of charged particles in the decay channel and
assign a systematic error of 0.8\% per track. The $\KS$ efficiency correction 
is taken from an independent study of the vertex-displacement dependence of the 
efficiency for inclusive samples of \KS mesons from the data and
from MC.  The overall correction for the topologies represented by our
decays is $0.971\pm0.030$.  For the six decays
with a primary \piz and the four with a \Kstarp\ or \rhop\ decaying to a
final state with an energetic \piz, we determine a correction from a sample of 
tau decays.  For these cases, the \piz\ efficiency is 6--11\% lower for data 
than MC.

\section{Fit Results}
\label{sec:results}

The branching fraction for each decay chain is obtained from
\beq
\calB = \frac{Y-Y_b}{\epsilon\prod{\calB_i}N_B}\,,
\eeq
where $Y$ is the yield of signal events from the fit, $Y_b$ is the fit
bias discussed in Sec.~\ref{sec:validation} and given in 
Table~\ref{tab:sumtab_all}, $\epsilon$ is the efficiency,
$\calB_i$ is the branching fraction for the $i$th unstable $B$ daughter
($\calB_i$ having been set to unity in the MC simulation), and $N_B$ is the 
number of produced \Bp\ or \Bz\ mesons.
The values of $\calB_i$ are taken from Particle Data Group world
averages~\cite{PDG2002}. The number of produced $B$ mesons is computed
with the assumption of equal production rates of charged and neutral \B\
pairs \cite{bbRpluszero}.

In Table~\ref{tab:restab_all}, we show the results of the final ML fits
to the on-peak data, with the yields for signal and \BB\ background,
where applicable.  The latter is often uncertain due to
the large correlation with the \qqbar\ background component, but this
uncertainty is not problematic because the correlation with signal is small.
We also show the efficiencies, daughter
branching-fraction products, and estimated
effective purity of the sample.  We report the statistical significance
for the individual decay chains and display the significance including
systematic uncertainties for the combined result in each channel.  The purity 
is the ratio of the signal yield to the effective background plus signal; we
estimate the denominator by taking the square of the uncertainty of the
signal yield as the sum of effective background plus signal.  Where the
signal yields are small the purity is not very meaningful, so we do not 
report the purity if it is below 10\%.  Branching fractions are given
for individual fits to each submode as well as the result of combining 
several submodes.  Since the latter procedure involves systematic as well as 
statistical errors, we defer the description to Sec.~\ref{sec:combine}. 
The final column in Table~\ref{tab:restab_all} gives the charge 
asymmetry (\acp ), as defined in Sec.~\ref{sec:intro}.

\begin{table*}[t]
\caption{Fitted event yields ($Y=$ signal yield, $Y_{\BB}=$ \BB\ yield), 
purity ($P$, see text), efficiency ($\epsilon$), 
daughter product branching fractions (in percent), 
significance ${\cal S}(\sigma)$ (which includes systematic errors),
fit branching fractions, 90\% C.L. upper limits,
and charge asymmetries.  Also shown are
the results of combining daughter decay chains where more than one 
contribute.  For the final branching fraction and 
charge asymmetry results, the systematic errors are also given.  }
\label{tab:restab_all}
\begin{center}
\vspace*{-0.3cm}
\begin{tabular}{lccrrrrccc}
    \dbline
Mode          		& $Y$ 		& $Y_{\BB}$     & $P$(\%)   & $\epsilon$~~ & $\prod\calB_i$~~	&~${\cal S}(\sigma$)& ${\cal B} (10^{-6})$ & UL $(10^{-6})$ & \acp  \\
\sgline
\bma{\etaKstp}		&    		&		&  	&    	&   		& \setaKstp~	&~\bma{\retaKstp}~& &~  \bma{\AetaKstp}\\
~~~\fetaggKstpKspip	& $46\pm12$     &  $25\pm15$	&  35~	& 24.0  &   $9.0~~$   	& 4.9 		& $22\pm6$	& &  $+0.03\pm0.24$\\
~~~\fetapppKstpKspip	& $27\pm8$      &		&  45~	& 17.1	&   $5.2~~$    	& 5.0 		& $33\pm10$	&&  $+0.46^{+0.24}_{-0.28}$\\
~~~\fetaggKstpKppiz	& $30\pm9$      &		&  45~	&  8.8	&  $13.1~~$   	& 5.7 		& $29\pm8$	& &  $-0.11\pm0.28$         \\
~~~\fetapppKstpKppiz	& $10\pm5$      &		&  43~	&  6.6	&   $7.5~~$    	& 3.2 		& $22\pm11$	&&  $+0.37^{+0.42}_{-0.51}$\\
\bma{\etaKstz}		&    		&		&   	&    	&   		& \setaKstz~	&~\bma{\retaKstz}~&&~  \bma{\AetaKstz}\\
~~~\fetaggKstz		& $125\pm16$	&   $5\pm19$	&  50~	& 24.4	&  $26.3~~$	& 10.1		& $20\pm3$	& &  $+0.12\pm0.13$\\
~~~\fetapppKstz		& $32\pm9$  	&		&  47~	& 16.5	&  $15.1~~$	& 5.0		& $14\pm4$	& &  $-0.39\pm0.25$\\
\bma{\etarhop}		&   		&		&  	&     	&  		& \setarhop	&\bma{\retarhop}&$<$\uletarhop &                         \\
~~~\fetaggrhop		& $32\pm15$     &  $-3\pm19$	&  14~	& 10.7	&  $39.4~~$	& 2.5		& $8\pm4$ 	& &                         \\
~~~\fetappprhop		& $21\pm11$	&   $3\pm11$	&  17~	&  8.6	&  $22.6~~$	& 2.4		& $12\pm6$	& &                        \\
\bma{\etarhoz}		& 		&		&  	& 	&  		& \setarhoz~	&\bma{\retarhoz}&$<$\uletarhoz &    			\\
~~~\fetaggrhoz		& $-18\pm18$	&  $67\pm38$	&$<$10~	& 27.1  &  $39.4~~$	& $-$~		&  $-2\pm2$     & &   			\\
~~~\fetappprhoz		&   $-2\pm4$    &  $26\pm10$	&$<$10~ & 18.2  &  $22.6~~$	& $-$~	 	&  $-1\pm1$     & &   			\\
\bma{\etapiz}		& 		&		&  	&  	&  		& \setapiz	&\bma{\retapiz}&$<$\uletapiz 	&   			\\
~~~\fetaggpiz		& $1\pm7$	&  $-2\pm9$	&$<$10~	& 19.3	&  $39.4~~$	& 0.3		& $0\pm1$	&&   \\
~~~\fetappppiz		& $8\pm7$	&  $-8\pm5$	&  15~	& 14.9	&  $22.6~~$	& 1.1		& $2\pm2$	&&   \\
\sgline
\bma{\etapKstp}		&     		&		&  	&    	&   		& \setapKstp	&\bma{\retapKstp}&$<$\uletapKstp &   \\
~~~\fetapeppKstpKspip	& $-8\pm4$	&  $29\pm11$	&$<$10~	& 17.5	&   $4.0~~$	& $-$~	 	& $-5\pm6$	&&   \\
~~~\fetaprgKstpKspip	& $16\pm9$	&  $17\pm12$	&  22~	& 13.5	&   $6.8~~$	& 1.7		& $~~15\pm11$	& & 	\\
~~~\fetapeppKstpKppiz	& $~3\pm3$	&		&  13~	&  7.0	&   $5.8~~$	& 1.7		& $~~8\pm7$	&  & 	\\
~~~\fetaprgKstpKppiz	& $~5\pm7$	&		&$<$10~	&  5.6	&   $9.8~~$	& 0.6		& $~~8\pm14$	&   &	\\
\bma{\etapKstz}		&   		&		&  	&    	&   		& \setapKstz	&\bma{\retapKstz}&$<$\uletapKstz &  \\
~~~\fetapeppggKstz	& $0\pm4$	&  $18\pm10$ 	&$<$10~	& 17.8	&  $11.6~~$	& 1.0 		& $2\pm2$      &&   \\
~~~\fetapfivepiKstz	& $~11\pm5$	&  $18\pm9$	&  47~	& 12.2	&   $6.7~~$	& 2.0		& $13\pm7$	&&   \\
~~~\fetaprgKstz		& $~~15\pm10$	&  $80\pm25$	&  17~	& 14.0	&  $19.7~~$	& 1.3		& $5\pm4$	& & 	\\
\bma{\etaprhop}		&   		&		&  	&    	&  		& \setaprhop	&\bma{\retaprhop}&$<$\uletaprhop &   \\
~~~\fetapepprhop	& $16\pm8$	&		&  25~	&  8.4	&  $17.5~~$	& 2.1		&$~11\pm6$      &&   \\
~~~\fetaprgrhop		& $48\pm23$	&  $61\pm100$	&$<$10~	&  6.5	&  $29.5~~$	& 1.7		&$~24\pm13$     & &  \\
\bma{\etaprhoz}		& $-1\pm4$	&  $53\pm21$	&$<$10~	& 19.7	&  $17.5~~$	& \setaprhoz	&\bma{\retaprhoz}&$<$\uletaprhoz &  \\
\bma{\etappiz}		&   		&  		&    	&       &  		& \setappiz	&\bma{\retappiz}&$<$\uletappiz &   \\
~~~\fetapepppiz		& $-2\pm3$	&  $-8\pm4$	&$<$10~	& 18.5	&  $17.5~~$	& 0.4		& $1\pm1$	&&   \\
~~~\fetaprgpiz		& $17\pm14$	& $-38\pm78$	&$<$10~	& 13.9	&  $29.5~~$	& 1.1		& $4\pm4$	& &  \\
\sgline
\bma{\omegapiz}		& $-9\pm8$	&   $9\pm18$	&$<$10~	& 15.9	&  $89.1~~$	&           	&\bma{\romegapiz}&$<$\ulomegapiz &  	\\
\sgline
\bma{\phipiz}		& $2\pm4$	&		&$<$10~	& 28.6	&  $49.2~~$	& \sphipiz	&\bma{\rphipiz}&$<$\ulphipiz 	&   	\\
\dbline
\end{tabular}
\end{center}
\end{table*}

The statistical error on the yield is given by the change in the central 
value when the quantity $-2\ln{\cal L}$ increases by
one unit.  The statistical significance is taken as the square root of the
difference between the value of $-2\ln{\cal L}$ for zero signal and the
value at its minimum. The 90\% C.L. upper limit quoted in 
Sec.~\ref{sec:conclusion}\ is the solution $\calB_{90}$ to the equation
\begin{equation}
\frac{\int_0^{\calB_{90}}{\calL}(b)db}{\int_0^\infty{\cal L}(b)db}=0.9\,,
\end{equation}
where ${\cal L}(b)$ is the value of the maximum likelihood for branching 
fraction $b$.

\begin{figure}[b!]
\vspace{0.5cm}
 \includegraphics[angle=0,width=\linewidth]{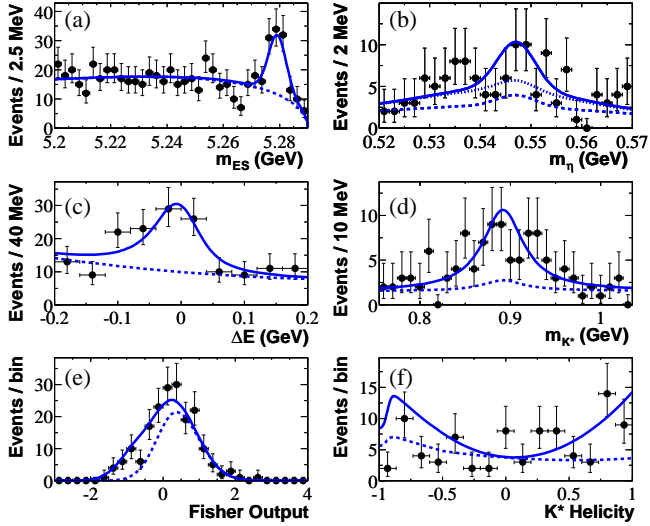}
 \caption{\label{fig:proj_etakstarp} Projections of the $B$-candidate
 discriminating variables for \etaKstp: (a) \mes; (b) $\eta$ candidate mass; 
(c) \DE; (d) \Kstp\ candidate mass; (e) Fisher discriminant output; and (f) 
\Kstp\ helicity.  See text for explanation of the points and curves.}
\end{figure}

\begin{figure}[htb!]
\vspace{0.5cm}
 \includegraphics[angle=0,width=\linewidth]{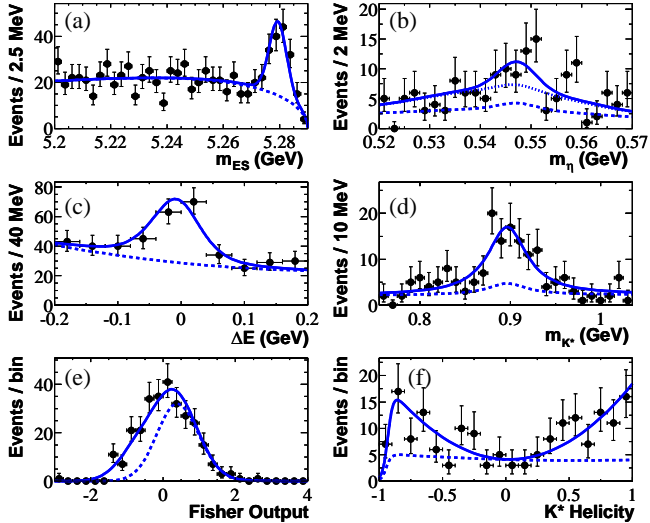}
 \caption{\label{fig:proj_etakstarz} Projections of the $B$ candidate
 discriminating variables for \etaKstz: (a) \mes; (b) $\eta$ candidate mass; 
(c) \DE; (d) \Kstz\ candidate mass; (e) Fisher discriminant output; and (f) 
\Kstz\ helicity.  See text for explanation of the points and curves.
}
\end{figure}

\begin{figure}[htb!]
\vspace{0.5cm}
 \includegraphics[angle=0,width=\linewidth]{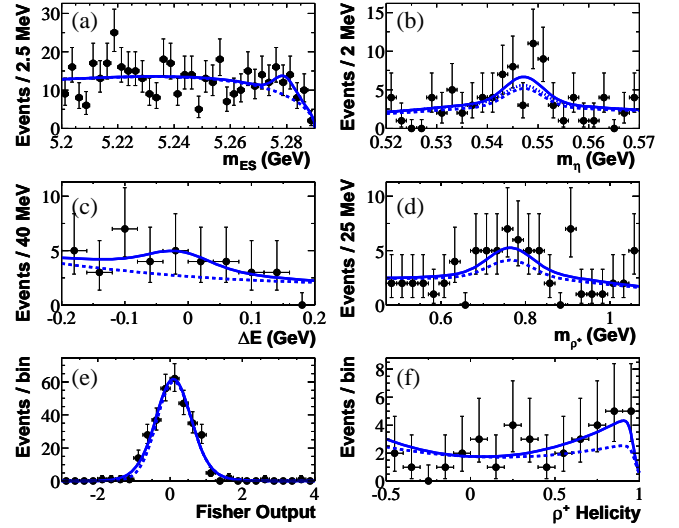}
 \caption{\label{fig:proj_etarhop} Projections of the $B$ candidate
 discriminating variables for \etarhop: (a) \mes; (b) $\eta$ candidate mass; 
(c) \DE; (d) \rhop\ candidate mass; (e) Fisher discriminant output; and (f) 
\rhop\ helicity.  See text for explanation of the points and curves.
}
\end{figure}

In Figs.~\ref{fig:proj_etakstarp}--\ref{fig:proj_etarhop} we show
projections of all fit discriminating variables for the \fetaKst\ and \fetarhop\ modes.
Points with errors represent data, solid curves the full fit functions, and
dashed curves the background functions.  Since the \etatogg\ and \etatoppp\
components have very different resolutions, for the $\eta$-candidate mass plots 
we indicate with a dashed curve the full fit without the \etatoppp\ signal
component.
We make these plots by selecting events with the ratio of signal to total 
likelihood (computed without the variable shown in the figure) exceeding a 
mode-dependent threshold that optimizes the expected sensitivity.  The selection
retains a fraction of the signal yield averaging about 70\% across the
decay sequences. 

\section{Systematic Uncertainties}
\label{sec:syst}

We itemize estimates of the various sources of systematic errors
important for these measurements.
Tables~\ref{tab:systtab_eta}, \ref{tab:systtab_etap}, and 
\ref{tab:systtab_pi0} show the results of
our evaluation of these uncertainties.  We tabulate separately the additive and
multiplicative uncertainties.  That is, we distinguish those errors affecting
the efficiency and total number of \BB\ events from those that concern the bias 
of the yield, since only the latter affect the significance of the result.
The two types of errors are comparable for modes with substantial yields
but the additive errors dominate when the yields are small.
Additionally we distinguish between those uncertainties
that are correlated among different daughter decays of the same mode
(C), and those that are uncorrelated (U).  This distinction is relevant
when multiple decay chains are combined (see Sec.~\ref{sec:combine}).
The final row of the table provides the total systematic error in the
branching fraction for each of the submodes.

\begin{table*}[htbp]
\caption{
Estimates of systematic uncertainties in the branching fraction for \etaKst\ and
\etarho\ decays.  We distinguish between additive and multiplicative errors as 
well as errors that are correlated (C) or uncorrelated (U) among the submodes.
}
\label{tab:systtab_eta}
\begin{center}
\begin{tabular}{lccrrccccrrccc}
\dbline
Quantity 		& \multicolumn{4}{c}{\fetaKstp}&\mdsp	& \multicolumn{2}{c}{\fetaKstz}&\mdsp	& \multicolumn{2}{c}{\fetarhop}\mdsp	& \multicolumn{2}{c}{\fetarhoz}&	\\
~~$\eta$ decay  	&\gaga    &\tpi     &\gaga~~   &\tpi~~   &\mdsp& \gaga	   & \tpi   &\mdsp	&\gaga~~    &\tpi~~ \mdsp&\gaga      & \tpi	  &	\\
~~~~\Kst , $\rho$ decay &$\Kz\pip$&$\Kz\pip$&$\Kp\piz$ &$\Kp\piz$&\mdsp&$\Kp\pim$ &$\Kp\pim$&\mdsp	&$\pip\piz$ &$\pip\piz$\mdsp&$\pip\pim$ &$\pip\pim$ &	\\
\sgline
Additive errors (events) \\
~~Fit yield (U)			& 2.1  &  0.7  &  0.4~ &  0.3~ & \mdsp&  2.2	&  2.3  &  \mdsp  &   2.3~ &  0.2~\mdsp&  5.2 	& 0.9  &  \\
~~Fit bias (U)			& 2.5  &  1.2  &  0.3~ &  0.3~ & \mdsp&  4.2	&  1.1  &  \mdsp  &   1.1~ &  0.9~\mdsp&  0.3 	& 0.7  &  \\
~~\BB\ background (U)   	& 1.1  &  0.3  &  0.6~ &  0.5~ & \mdsp&  0.9	&  0.3 	&  \mdsp  &   0.7~ &  0.4~\mdsp&  6.5	& 1.2  &  \\\sgline                                                                 
Total additive (events) 	& 3.4  &  1.4  &  0.6~ &  0.5~ & \mdsp&  4.8	&  2.6	&  \mdsp  &   2.6~ &  1.0~\mdsp&  8.1	& 1.6  &  \\
\sgline 
Multiplicative errors (\%)~~~~ \\
~~Tracking eff/qual (C) 	& 0.8  &  2.4  &  0.8~ &  2.4~ & \mdsp&  1.6	&  3.2  &  \mdsp  &   0.8~ &  2.4~\mdsp& 1.6  	&  3.2 &  \\
~~\KS\ efficiency (C)   	& 4.0  &  4.0  &       &       & \mdsp&     	&       &  \mdsp  &        &      \mdsp&      	&      &  \\
~~Track multiplicity (C)	& 1.0  &  1.0  &  1.0~ &  1.0~ & \mdsp&  1.0	&  1.0  &  \mdsp  &   1.0~ &  1.0~\mdsp& 1.0  	&  1.0 &  \\
~~\piz/ $\gamma$ eff (C)	& 5.1  &  5.1  & 10.3~ & 10.3~ & \mdsp&  5.1	&  5.1  &  \mdsp  &  10.3~ & 10.3~\mdsp& 5.1  	&  5.1 &  \\
~~Number \BB\ (C)       	& 1.1  &  1.1  &  1.1~ &  1.1~ & \mdsp&  1.1	&  1.1  &  \mdsp  &   1.1~ &  1.1~\mdsp& 1.1  	&  1.1 &  \\
~~Branching fractions (U)	& 0.7  &  1.8  &  0.7~ &  1.8~ & \mdsp&  0.7	&  1.8  &  \mdsp  &   0.7~ &  1.8~\mdsp& 0.7  	&  1.8 &  \\
~~MC statistics (U)     	& 0.9  &  1.1  &  1.5~ &  1.8~ & \mdsp&  0.9	&  1.1  &  \mdsp  &   1.3~ &  1.7~\mdsp& 0.8  	&  1.0 &  \\
~~\costhr\ (C)          	& 0.5  &  0.5  &  0.5~ &  0.5~ & \mdsp&  0.5	&  0.5  &  \mdsp  &   1.0~ &  0.5~\mdsp& 2.3  	&  1.0 &  \\
~~PID (C)               	&      &       &  1.0~ &  1.0~ & \mdsp&  1.0	&  1.0  &  \mdsp  &        &      \mdsp&     	&      &  \\
\sgline
Total multiplicative (\%)	& 6.8  &  7.4  & 10.6~ & 11.0~ & \mdsp&  5.8 	&  6.6  &  \mdsp  &  10.6~ & 11.0~\mdsp&  6.1  & 6.6  &  \\
\sgline
Total $\sigma$ \lbrack\bfemsix\rbrack	& 2.2  &  2.9  &  3.6~ &  2.6~ & \mdsp&  1.4	&  1.5  &  \mdsp  &   1.0~ &  1.5~\mdsp&  0.9 	& 0.4  &  \\
\dbline
\end{tabular}
\end{center}
\end{table*}

\begin{table*}[htbp]
\caption{
Estimates of systematic uncertainties in the branching fraction for \etapKst\ 
and \etaprho\ decays.  The notation is the same as for Table \ref{tab:systtab_eta}.
}
\label{tab:systtab_etap}
\begin{center}
\begin{tabular}{lccrccccccrrcc}
\dbline
Quantity & \multicolumn{4}{c}{\fetapKstp}&\mdsp	& \multicolumn{3}{c}{\fetapKstz}&\mdsp	& \multicolumn{2}{c}{\fetaprhop}	&\mdsp	& {\fetaprhoz}\\
~~$\etapr$ decay	&\epp	&\rg	&\epp~	&\rg	&\mdsp	&\epp	&\fivepi &\rg		&\mdsp	&	\epp~	&\rg~~	&\mdsp	& \epp		\\
~~~~\Kst , $\rho$ decay &$\Kz\pip$&$\Kz\pip$&$\Kp\piz$&$\Kp\piz$&\mdsp&$\Kp\pim$ &$\Kp\pim$&$\Kp\pim$&\mdsp	&$\pip\piz$ &$\pip\piz$	&\mdsp&$\pip\pim$	\\
\sgline
Additive errors (events) \\
~~Fit yield (U)           &  0.9  &  1.3  &  0.2~ &  0.8  & \mdsp&  0.5  &  2.8  &  0.8 & \mdsp&  0.8~ &  3.1~ & \mdsp&  1.0     \\
~~Fit bias (U) 		  &  2.3  &  1.9  &  0.3~ &  0.4  & \mdsp&  1.7  &  0.8  &  2.1 & \mdsp&  1.2~ &  4.3~ & \mdsp&  1.9     \\
~~\BB\ background (U)     &  1.3  &  0.9  &  0.5~ &  0.5  & \mdsp&  2.5  &  0.5  &  1.4 & \mdsp&  1.2~ & 10.0~ & \mdsp&  1.8     \\
\sgline                                                                                                                          
Total additive (events)   &  2.8  &  2.5  &  0.6~ &  1.0  & \mdsp&  3.1  &  3.0  &  2.6 & \mdsp&  1.9~ & 11.3~ & \mdsp&  2.8     \\
\sgline
Multiplicative errors (\%)~~~~ \\
~~Tracking eff/qual (C)   &  2.4  &  2.4  &  2.4~ &  2.4  & \mdsp&  3.2  &  4.8  &  3.2 & \mdsp&  2.4~ &  2.4~ & \mdsp&  3.2     \\
~~\KS\ efficiency (C)     &  4.0  &  4.0  &       &       & \mdsp&       &       &      & \mdsp&       &       & \mdsp&           \\
~~Track multiplicity (C)  &  1.0  &  1.0  &  1.0~ &  1.0  & \mdsp&  1.0  &  1.0  &  1.0 & \mdsp&  1.0~ &  1.0~ & \mdsp&  1.0     \\
~~\piz / $\gamma$ eff (C) &  5.4  &  2.5  & 10.4~ &  7.6  & \mdsp&  5.4  &  5.4  &  2.5 & \mdsp& 10.4~ &  7.6~ & \mdsp&  5.4     \\
~~Number \BB\ (C)         &  1.1  &  1.1  &  1.1~ &  1.1  & \mdsp&  1.1  &  1.1  &  1.1 & \mdsp&  1.1~ &  1.1~ & \mdsp&  1.1     \\
~~Branching fractions (U) &  3.4  &  3.4  &  3.4~ &  3.4  & \mdsp&  3.4  &  3.4  &  3.4 & \mdsp&  3.4~ &  3.4~ & \mdsp&  3.4     \\
~~MC statistics (U)       &  1.0  &  1.2  &  1.7~ &  1.9  & \mdsp&  1.1  &  1.1  &  1.1 & \mdsp&  1.5~ &  1.8~ & \mdsp&  0.9     \\
~~\costhr\ (C)            &  0.5  &  1.8  &  0.5~ &  1.8  & \mdsp&  0.5  &  1.4  &  1.4 & \mdsp&  0.5~ &  3.0~ & \mdsp&  0.5     \\
~~PID (C)                 &       &       &  1.0~ &  1.0  & \mdsp&  1.0  &  1.0  &  1.0 & \mdsp&       &       & \mdsp&           \\
\sgline
Total multiplicative (\%) &  8.1  & 6.8   & 11.5~ &  9.2  & \mdsp&  7.5	 &  8.4  &  5.4 & \mdsp& 11.4~ &  9.5~ & \mdsp&  7.4     \\
\sgline
Total $\sigma$ \lbrack\bfemsix\rbrack& 4.5 &  3.2  &  1.9~ &  2.1  & \mdsp&  1.7	 &  4.2  & 1.1 & \mdsp&  1.9~ &  6.9~  & \mdsp&  0.9    \\
\dbline
\end{tabular}
\end{center}
\end{table*}

\begin{table}[htbp]
\caption{
Estimates of systematic uncertainties in the branching fraction for the decays 
\etapiz, \etappiz, \omegapiz and \phipiz.
The notation is the same as for Table \ref{tab:systtab_eta}.
}
\label{tab:systtab_pi0}
\begin{minipage}{\linewidth}
\begin{tabular}{lrrrrrc}
\dbline
Quantity 	& \multicolumn{2}{c}{\fetapiz}	& \multicolumn{2}{c}{\fetappiz}	& \fomegapiz~&~\fphipiz~\\
~~~~$\eta^{(\prime)}$ decay &$\gaga$	&$\tpi$~&~$\epp$	&$\rg$   &		&		\\
\sgline
Add. err. (evts.) \\
~~Fit yield (U)     	&  1.0	&  1.3	 	&  1.2 	&  2.8	&  1.4 	&  0.8\\
~~Fit bias (U) 		&  1.0	&  0.3	  	&  2.0	&  1.3	&  1.5 	&  0.7\\
~~\BB\ Bkg (U)     	&  1.0	&  1.0		&  1.0	&  1.0	&  1.0 	&  1.0\\
\sgline                       
Total add. (evts.) 	&  1.7	&  1.7	      	&  2.5	&  3.2	&  2.3 	&  1.5\\
\sgline                        
Mult. err. (\%) \\
~~Track eff. (C)   	&  0.0	&  1.6	  	&  1.6	&  1.6	&  1.6	&  1.6\\
~~Multiplicity (C)  	&  1.0	&  1.0	  	&  1.0	&  1.0	&  1.0	&  1.0\\
~~\piz / $\gamma$ eff (C)& 14.9 & 11.5	  	& 13.1	&  9.1	&~~11.7	&  7.9\\
~~Number \BB\ (C)       	&  1.1	&  1.1	 	&  1.1	&  1.1	&  1.1	&  1.1\\
~~Br. frac. (U) 	&  0.8	&  1.8	  	&  3.4	&  3.4	&  0.8	&  1.4\\
~~MC stats. (U)       	&  1.2	&  1.3	  	&  1.2	&  0.5	&  1.2	&  0.9\\
~~\costhr\ (C)    	&  0.5	&  0.5	  	&  0.5	&  1.3	&  0.6 	&  0.5\\
\sgline
Total mult. (\%)        & 15.1	& 11.9       	& 13.8	& 10.1	& 12.0  &  8.4\\
\sgline
Total $\sigma$ \lbrack\bfemsix\rbrack&0.3&  0.6  	&  0.9	&  1.0  &  0.2 	&  0.1\\
\dbline
\end{tabular}
\end{minipage}
\end{table}

\subsection{Additive systematic errors}

\noindent\underline{Fit yield} (U):  
Uncertainties due to imprecise knowledge of the background PDF parameters are 
included in the statistical errors since the main parameters are allowed to 
vary in the nominal fits.  We have investigated the small correlations among
background parameters and find these to have a negligible effect on
signal yields.  We include the uncertainty
for the signal PDF parameters by determining the yield variations as
individual parameters are varied by uncertainties determined from fits to 
independent control samples (see Sec.~\ref{sec:control}).

\noindent\underline{Fit bias} (U):
This uncertainty is taken from the validation procedure described in
Sec.~\ref{sec:validation}.  We combine in quadrature terms, in order of
relative importance, from (a) the
positive bias (due to parameter correlations), (b) the negative bias for 
small event yields, (c) a small contribution from the modeling of the
combinatorial component in signal, and (d) the statistical uncertainty in the 
determination of the bias.  The first uncertainty (a) is taken to be 
one half of the positive bias, and the second (b) to be one half of the 
difference between the peak 
and mean yields of the ensemble distributions.  Contribution (c)
is small for all modes; we determine it using a comparison of 
Monte Carlo and data for the $\Bm\ra\pim\Dz$ control sample.

\noindent\underline{\BB\ background} (U): 
The \BB\ background component, included in the fit for most decay chains, 
accounts for most uncertainties from \BB\ background.  We assign an
additional uncertainty to account for modeling of this background.
For the high-background \fetaprhop\ decay this involves explicit
variation of the model. For the other modes it is taken to be 50\%
of the difference in the signal yields when background is varied by its
uncertainty (100\% of the estimated effect when a \BB\ background component
is not included in the fit) and a contribution to account for 
uncertainty in the effect of the $b\ra c$ background.

\subsection{Multiplicative systematic errors}

\noindent\underline{Track finding/efficiency} (C): 
As described in Sec.~\ref{sec:eff}, we assign a systematic error of
0.8\% for each track (except for those from \kzs\ decays - see below).

\noindent\underline{\KS\ reconstruction efficiency} (C):
The $\KS$ efficiency systematic uncertainty is taken from the study described in
Sec.~\ref{sec:eff} with the addition of a contribution for
reconstruction of the daughter charged tracks, giving a total uncertainty of 4\%\
for decays with a \KS\ in the final state.

\noindent\underline{Track multiplicity} (C): 
The inefficiency of the preselection requirements for the number of
tracks in the event is a few percent. We estimate an uncertainty of 1\%
from the uncertainty in the low-multiplicity tail of the \B\ decay
model.

\noindent\underline{$\gamma$, \piz, $\eta_{\gamma\gamma}$ reconstruction 
efficiency} (C): This uncertainty is estimated to be 2.5\%/photon from a study of
tau decays to modes with \piz's.  For \piz's with energy greater than 1 \gev, 
there is an additional contribution to the uncertainty due to the overlap of 
the two showers, also evaluated from tau decays.

\noindent\underline{Luminosity, $B$ counting} (C):  
From a sample of $e^+e^-\to\mu^+\mu^-$ decays, we estimate the uncertainty on 
the number of produced \BB\ pairs to be 1.1\%.

\noindent\underline{Branching fractions of decay chain daughters} (U): 
This is simply taken as the uncertainty on the daughter particle
branching fractions from Ref.~\cite{PDG2002}.

\noindent\underline{MC statistics} (U): 
The uncertainty due to finite signal MC sample sizes (typically 40,000
generated events) is given in the table. 

\noindent\underline{Event shape requirements} (C): 
The uncertainties in the Fisher distribution \xf\ are included in the fit yield 
systematic variation (see below).  Uncertainties due to the \costhr\ requirement
are estimated to be one-half of the difference between the
observed signal MC efficiency for the \costhr\ requirement used for each
analysis and the expectation for a flat distribution.

\noindent\underline{PID} (C):  
The uncertainties due to PID vetoes are negligible.  For analyses with a 
charged kaon, we estimate from independent samples an average efficiency
uncertainty of 1.0\%.

\subsection{Charge asymmetry systematic errors}
\label{sec:chgasymsyst}

For the \etaKst\ analyses, the charged $K$ 
used to define the asymmetry has a broad momentum spectrum.  Auxiliary 
tracking studies place a stringent bound on detector charge-asymmetry
effects at all momenta.  Such tracking and PID systematic effects were
studied in detail for the analysis of $B\ra\phi \Kstar$~\cite{phiKstarPRL}.
We assign the same 2\% systematic uncertainty for \acp\ that was
determined in that study.  In addition, we observe that the charge
asymmetry of the continuum background is consistent with zero in all
cases with a combined uncertainty below 1\%.
Finally we have measured the charge asymmetry for a control
sample of $B^-\ra D^0\rho^-$ decays and find the result to be
consistent with zero asymmetry, as expected.

\section{\boldmath Combined Results}
\label{sec:combine}

To obtain the final results, we combine the branching fraction and charge
asymmetry measurements from the individual daughter decay chains.  The
joint likelihood is given by the product, or equivalently $-2\ln\calL$
is given by the sum, of contributions from the submodes.  The
statistical contribution comes directly from the likelihood fit, which
reflects the non-Gaussian uncertainty associated with small numbers of events.
Before combining, we convolve each statistical $\calL$ with a Gaussian
function representing the part of the systematic error that is uncorrelated 
among the submodes.  The $-2\ln\calL$ distributions without systematic 
uncertainties give the combined statistical errors, while the distributions 
including correlated systematic uncertainties, give the total statistical and 
systematic errors.

The resulting branching fractions and charge asymmetries are included
in Table~\ref{tab:restab_all}, where the significance includes
systematic uncertainties.

\section{Discussion}
\label{sec:discussion}

More than six years have passed since the first report of a very large 
branching fraction for the decay \etapK, published in Ref.~\cite{CLEOetapobs}.
While it was expected \cite{Lipkin} that the branching fraction for 
this decay and \etaKst\ would be
relatively large and \etaK\ and \etapKst\ would be much smaller, most
theoretical calculations could not account for a branching fraction as
large as was measured.  The experimental situation with \etapK\ has 
remained largely the same even with quite precise new measurements; 
see for example Ref.~\cite{etaprPRL}.  The results presented in this paper
complete the measurement of the four $(\eta,\etapr)(K,\Kstar)$ final
states with a sensitivity in the branching fraction of a few times $10^{-6}$. 
The \etaKst\ decays are found to have rather large branching fractions 
as expected and as first seen by CLEO \cite{CLEOetapr}.  \babar\ has
recently observed \etaKp\ for the first time \cite{etaomegaPRL} and 
finds the expected small branching fraction.
We find no significant signal for \etapKst, and the 90\% C.L. upper limit 
is not yet precise enough to determine whether a flavor-singlet component 
is present for this decay, though we do restrict the size of such a contribution.
Such a singlet component (see Fig.~\ref{fig:feyn_etakstar}d) has been
proposed as a partial explanation for the large rate for \etapK\ by many
authors, though with the restrictive limits for \etapKst, this now seems
unlikely to play a significant role \cite{etapkst}.

We also have evidence for the decay \etarhop\ with a significance of
\setarhop$\sigma$.  We find no other significant signals and calculate upper 
limits for \etaprhop\ and all of the neutral $B$ decays with a $\rho$ or 
$\piz$ meson.  This pattern is as expected since the penguin contribution 
in these decays is CKM suppressed and there is no external tree diagram
for the \Bz\ decays.

For the decays where we find significant signals, we also measure the charge 
asymmetry, which we find to be consistent with zero for all of these decays.  
These measurements are in agreement with the theoretical expectations discussed
in Sec.~\ref{sec:intro}\ and rule out substantial portions of the
physical region.

\section{Summary of results}
\label{sec:conclusion}

We report measurements of branching fractions and charge asymmetries
for $B$-meson decays to $\eta$ or \etapr\ with a $K^*$, $\rho$, or
$\piz\!$ as well as those channels with an \om\ or $\phi$ and a \piz .
We find signals with high statistical significance in the \etaKst\
channels. We have evidence for the decay \etarhop\  
(with significance $\setarhop\sigma$), which has
not been seen previously.  For branching fractions with significance less 
than four standard deviations, we quote both central values with errors 
and 90\% C.L. upper limits.  The observed values in the $\eta$ channels are
\begin{eqnarray*}
\BetaKstp &=& \RetaKstp\,, \\
\BetaKstz &=& \RetaKstz\,, \\
\Betarhop &=& \Retarhop    \\
	  &<& \ULetarhop\,,\\
\Betarhoz &=& \Retarhoz    \\
	  &<& \ULetarhoz\,,\\
\Betapiz  &=& \Retapiz     \\
	  &<& \ULetapiz\,.
\end{eqnarray*}
For the \etapr\ channels, we find
\begin{eqnarray*}
\BetapKstp &=& \RetapKstp    \\
	   &<& \ULetapKstp\,, \\
\BetapKstz &=& \RetapKstz    \\
	   &<& \ULetapKstz\,, \\
\Betaprhop &=& \Retaprhop     \\
	   &<& \ULetaprhop\,, \\
\Betaprhoz &=& \Retaprhoz     \\ 
	   &<& \ULetaprhoz\,, \\
\Betappiz  &=& \Retappiz      \\
	   &<& \ULetappiz\,.
\end{eqnarray*}
In the modes with a vector meson and a \piz, we observe
\begin{eqnarray*}
\Bomegapiz &=& \Romegapiz\,,  \\
	   &<& \ULomegapiz\,, \\
\Bphipiz   &=& \Rphipiz\,,    \\
	   &<& \ULphipiz\,.
\end{eqnarray*}
The results for \omegapiz\ supersede the previous \babar\ measurement of for 
this channel~\cite{BABARetapOmega}. All of these results are substantially
more precise than previous measurements from CLEO \cite{CLEOetapr}.

For the modes with significant signals, we measure the charge asymmetries
\begin{eqnarray*}
\acp(\fetaKstp) &=& \AetaKstp\,, \\
\acp(\fetaKstz) &=& \AetaKstz\,.
\end{eqnarray*}

\section{Acknowledgments}
\label{sec:Acknowledgments}

We thank Michael Gronau and Jon Rosner for useful discussions.
We are grateful for the 
extraordinary contributions of our \pep2\ colleagues in
achieving the excellent luminosity and machine conditions
that have made this work possible.
The success of this project also relies critically on the 
expertise and dedication of the computing organizations that 
support \babar.
We wish to acknowledge support from the University of Colorado Undergraduate 
Research Opportunities Program.
The collaborating institutions wish to thank 
SLAC for its support and the kind hospitality extended to them. 
This work is supported by the
US Department of Energy
and National Science Foundation, the
Natural Sciences and Engineering Research Council (Canada),
Institute of High Energy Physics (China), the
Commissariat \`a l'Energie Atomique and
Institut National de Physique Nucl\'eaire et de Physique des Particules
(France), the
Bundesministerium f\"ur Bildung und Forschung and
Deutsche Forschungsgemeinschaft
(Germany), the
Istituto Nazionale di Fisica Nucleare (Italy),
the Foundation for Fundamental Research on Matter (The Netherlands),
the Research Council of Norway, the
Ministry of Science and Technology of the Russian Federation, and the
Particle Physics and Astronomy Research Council (United Kingdom). 
Individuals have received support from 
the A. P. Sloan Foundation, 
the Research Corporation,
and the Alexander von Humboldt Foundation.


\begin{thebibliography}{99}
\bibitem{CC}
Except as noted otherwise, we use a particle name to denote either member of 
a charge-conjugate pair.
\end{thebibliography}
\end{document}